\documentclass[twocolumn,tighten,twocolappendix]{aastex63}

\usepackage{ulem}

\graphicspath{{./}{figures/}}

\received{November, 11, 2019}
\revised{June, 3, 2020}
\accepted{June, 22, 2020}
\submitjournal{ApJ}

\shorttitle{Circumstellar Disk Properties}
\shortauthors{Hirano et al.}

\newcommand{\msun}{M_\odot}

\newcommand{\nh}{n_{\rm H}}
\newcommand{\cc}{{\rm cm^{-3}}}
\newcommand{\kms}{{\rm km\,s^{-1}}}
\newcommand{\vrad}{v_{\rm rad}}
\newcommand{\vect}[1]{\mbox{\boldmath$#1$}}

\renewcommand{\labelenumi}{(\Alph{enumi})}

\begin{document}

\title{The Effect of Misalignment between Rotation Axis and Magnetic Field on Circumstellar Disk} Properties

\correspondingauthor{Shingo Hirano}
\email{hirano.shingo.821@m.kyushu-u.ac.jp}

\author[0000-0002-4317-767X]{Shingo Hirano}
\affiliation{Department of Earth and Planetary Sciences, Faculty of Sciences, Kyushu University, Fukuoka 819-0395, Japan}

\author{Yusuke Tsukamoto}
\affiliation{Department of Earth and Space Science, Graduate Schools of Science and Engineering, Kagoshima University,\\ Kagoshima 890-0065, Japan}

\author{Shantanu Basu}
\affiliation{Department of Physics and Astronomy, The University of Western Ontario, London, ON N6A 3K7, Canada}

\author{Masahiro N. Machida}
\affiliation{Department of Earth and Planetary Sciences, Faculty of Sciences, Kyushu University, Fukuoka, Fukuoka 819-0395, Japan}
\affiliation{Department of Physics and Astronomy, The University of Western Ontario, London, ON N6A 3K7, Canada}

\begin{abstract}
The formation of circumstellar disks is investigated using three-dimensional resistive magnetohydrodynamic simulations, in which the initial prestellar cloud has a misaligned rotation axis with respect to the magnetic field. 
We examine the effects of (i) the initial angle difference between the global magnetic field and the cloud rotation axis ($\theta_0$) and (ii) the ratio of the thermal to gravitational energy ($\alpha_0$).
We study $16$ models in total and calculate the cloud evolution until $\sim \! 5000$\,yr after protostar formation. 
Our simulation results indicate that an initial non-zero $\theta_0$ ($> 0$) promotes the disk formation but tends to suppress the outflow driving, for models that are moderately gravitationally unstable, $\alpha_0 \lesssim 1$. 
In these models, a large-sized rotationally-supported disk forms and a weak outflow appears, in contrast to a smaller disk and strong outflow in the aligned case ($\theta_0 = 0$).
Furthermore, we find that when the initial cloud is highly unstable with small $\alpha_0$, the initial angle difference $\theta_0$ does not significantly affect the disk formation and outflow driving.
\end{abstract}


\section{Introduction}
\label{sec:intro}

Circumstellar disks are a by-product of star formation and the hosts of planet formation. 
Thus, the formation and evolution of circumstellar disks should be clarified in order to understand both the star and planet formation processes.
Recent ALMA observations show that circumstellar disks exist even in a very early phase of star formation: the so-called Class 0 stage \citep[e.g.,][]{sakai14,ohashi14,lefloch15,plunkett15, ching16,tokuda16,aso17,lee17,lee18}.
At this stage, the typical disk size is considered to be as small as $\lesssim\!10$\,au \citep{yen17}, although large-sized disks are sometimes observed \citep[e.g.,][]{hara13,okoda18}.
Very recently, the disk structures during the later Class I and II stages of star formation were also revealed by ALMA \citep[e.g.,][]{aso15,bjerkeli16,perez16,alves17}.
The DSHARP project showed beautiful views of disks around Class II  pre-main-sequence stars, in which various morphologies of ring, gap, and spiral patterns were clearly shown \citep[][and references therein]{andrews18}.
Understanding the early phase of the disk evolution (i.e., Class 0 and early Class I stages) is important to better clarify the late phase of the disk (i.e., late Class I and Class II stages).

In addition to the progress in observations, numerical simulations have greatly contributed to understanding disk formation. 
Specifically, since the studies reported by \citet{allen03} and \citet{mellon08},  the magnetic braking (problem) has been intensively discussed in the context of disk formation \citep{li14}.
Based on early simulations, some researchers claimed that the circumstellar disk cannot be formed in the very early phase of star formation, because the angular momentum is excessively removed from the disk-forming region by magnetic braking (the so-called magnetic braking catastrophe; \citealt{mellon09} and \citealt{li13}).

It turns out that non-ideal magnetohydrodynamical (MHD) effects (Ohmic dissipation, ambipolar diffusion, and Hall effect) are considerably important to investigate the disk evolution, because a nascent disk forms in a magnetically inactive region where the magnetic field is weakened by Ohmic dissipation and ambipolar diffusion \citep[][see also reviews by \citealt{tsukamoto16} and \citealt{wurster18}]{dapp10,dapp12,tomida15,masson16,tsukamoto18}.
The diffusion rates of Ohmic dissipation and ambipolar diffusion strongly depend on the star-forming environment.
The amount of charged particles and dust grains and the strength of cosmic rays determine the coefficients of magnetic diffusion rates and therefore the resultant disk size \citep{marchand16,zhao16,zhao18,wurster18b,koga19}.

The Hall effect is also important for disk formation because it can generate toroidal magnetic fields from poloidal fields in the collapsing cloud. This strengthens or weakens the magnetic braking depending on the relative direction of the magnetic field and the rotation axis \citep[magnetic field parallel or anti-parallel to rotation axis; e.g.,][]{krasnopolsky11,tsukamoto15b, tsukamoto17}.
In addition to the non-ideal MHD effects, some studies showed that turbulence in the star-forming core can play a role in the formation and evolution of circumstellar disks even when the prestellar cloud has a strong magnetic field \citep{santos13,seifried12,seifried13}.

Several scenarios have been offered to explain the observed range of disk sizes and masses.
The disk size in the early phase (or Class 0 stage) could vary for prestellar clouds with different initial magnetic field strengths and rotation rates \citep[e.g.,][]{masson16}.
Fundamentally, the disk size is determined by the balance between the angular momentum falling into the central region and that transported outward by the magnetic effects \citep{machida11}.
An initially rapid rotation can produce a large-sized disk and a strong magnetic field can suppress the disk size growth.  
The initial cloud shape and the distributions of gas density, angular velocity, and magnetic field would also change the disk size \citep[see \S\,1 of][]{gray18}.
Note, however, that \citet{hennebelle16} claimed that the disk size is determined only by the magnetic field strength of the disk and the ambipolar diffusion coefficient in their analytic study.

Most three-dimensional (3D) MHD simulations have, for simplicity, studied aligned rotators, i.e., the rotation axis of the prestellar cloud is parallel to the global magnetic field \citep[e.g.,][]{tomisaka02,machida04,banerjee06,tomida13,tsukamoto15}, and a comprehensive picture has emerged in this scenario \citep[see][]{inutsuka12}.
For the first time, \cite{matsumoto04} and \cite{matsumoto06} investigated the  misalignment effect, i.e., when the rotation axis of the initial cloud is misaligned with the global magnetic field, with their three dimensional MHD simulations. 
Then, \citet{hennebelle09} pointed out that the disk formation process is significantly affected by misalignment.
Since then, many researchers have discussed the importance of misalignment in the process of disk formation \citep[e.g.,][and see \S\ref{sec:past} for details]{joos12}.
The classical effects of magnetic braking will tend to bring a prestellar core into alignment, since angular momentum that is perpendicular to the global magnetic field is lost more efficiently than the parallel component \citep{mouschovias79,mouschovias80}.
On the other hand, the presence of turbulence could produce an angle difference (i.e., misalignment) between the magnetic field and rotation vector \citep{joos13,krumholz13,matsumoto17}.
Thus, a misalignment seems to be a natural consequence in clouds with both turbulence and magnetic field.

This study focuses on the effect of misalignment on the disk formation.
Using a resistive MHD simulation with a sink cell, we investigate the evolution of star-forming cloud cores in which the rotation axis of the initial cloud is not aligned with the global magnetic field.
The initial state is still not highly turbulent, and maintains other symmetries even though there is misalignment (for details, see \S\ref{sec:past}). 
We believe that disk formation should be investigated in this still idealized setting in order to proceed in a step-by-step manner, rather than trying to investigate it a very complicated environment such as a highly turbulent cloud. 
Finally, we comment on the relation  between our previous  \citep{machida20} and present studies, in which the initial conditions (or initial prestellar clouds) are almost the same.
\citet{machida20} focused on the small scale structures only for $\sim\!500$\,yr after protostar formation. They resolved a protostar without a sink cell and qualitatively compared the directions of the jet, disk and local magnetic field with those of observations. 
In this study, we use a sink cell and focus on the long-term ($\sim\!5000$\,yr) evolution of the circumstellar disk and outflow and the efficiency of disk formation.

Our paper is structured as follows. 
We summarize the past misalignment studies in \S2. 
Numerical settings and initial conditions of our study are described in \S3.
Our simulation results are presented in \S4. 
We discuss the misalignment and disk formation in \S5, and compare simulations with observations in \S6.
We summarize our results in \S7.

\section{Past Studies}
\label{sec:past}

We first comment on analytic studies, which had been performed mainly by Mouschovias and his collaborators \citep[e.g.,][]{mouschovias79, mouschovias80, mouschovias80b, mouschovias85, mouschovias86}. 
They investigated the efficiency of magnetic braking in different configurations of the magnetic field. 
It should be noted that although magnetic braking from a collapsing cloud core was considered in some such studies during 1970s and 1990s \citep[see also][]{basu94,basu95a,basu95b}, they did not specifically focus on the circumstellar disk formation.

\begin{figure}[t]
  \begin{center}
    \includegraphics[width=1.0\columnwidth]{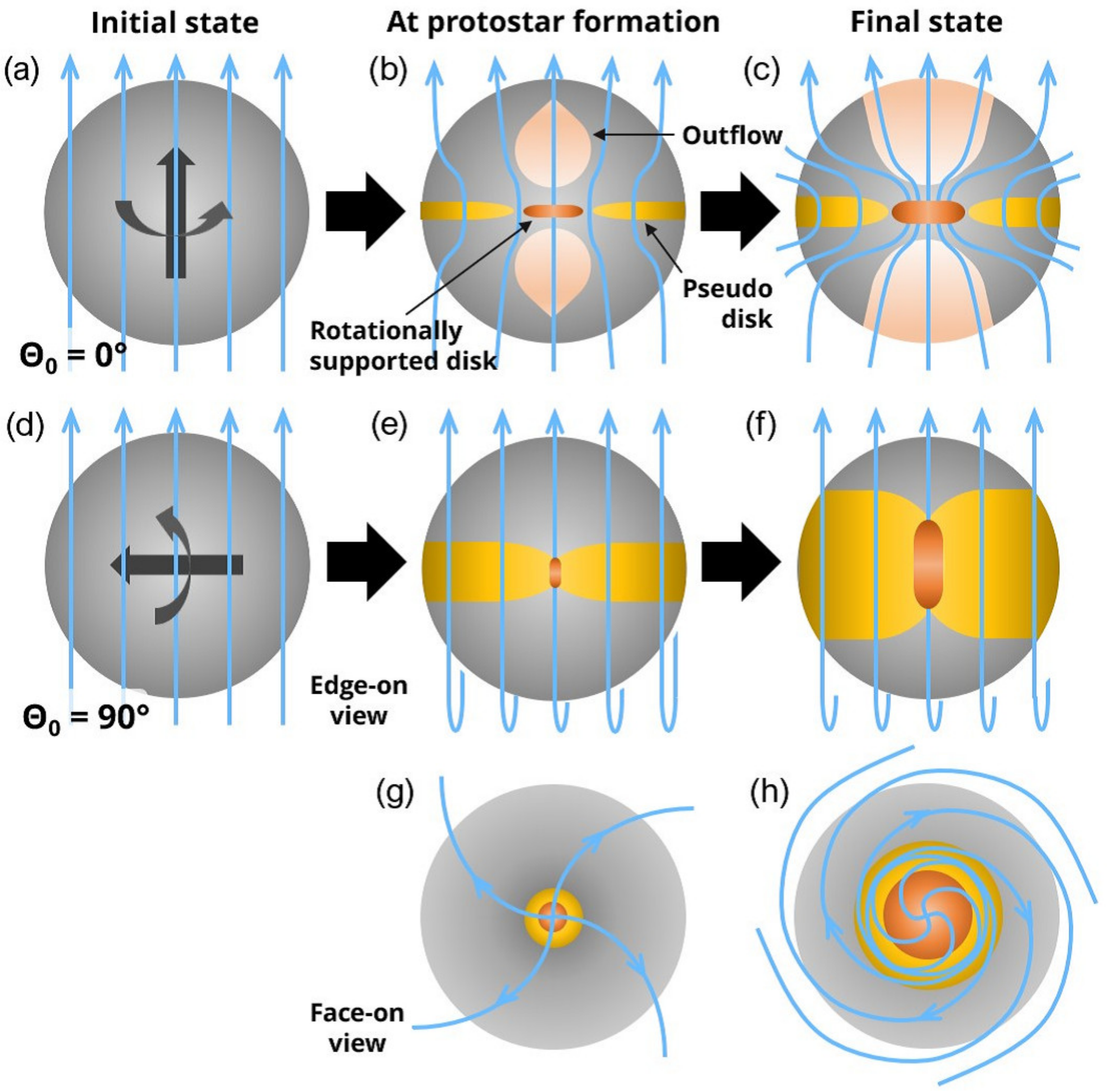}
  \end{center}
\caption{
Schematic view of the time sequence of a star-forming cloud for the aligned ($\theta_0 = 0^\circ$; panels a--c) and perpendicular ($\theta = 90^\circ$; d--h) cases.
Panels (a)--(f) are the edge on view, while panels (g) and (h) correspond to the face-on view of panels (e) and (f), respectively.
The large black arrow indicates time evolution.
Magnetic field lines are indicated by blue lines.
Gray, yellow and orange regions indicate the whole star-forming cloud, pseudodisk, and rotationally-supported disk, respectively. 
In panels (a) and (d), the large black arrow in the sphere corresponds to the rotation axis.
In panels (b) and (c), the outflow is represented by a red color.
}
\label{f1}
\end{figure}

\citet{mouschovias80} showed that magnetic braking is more efficient in the perpendicular configuration than in the aligned uniform configuration, because a strong magnetic tension force brakes the disk (Figure~\ref{f1}h).
On the other hand, magnetic braking in the aligned fan-shaped configuration can be more efficient than in the perpendicular configuration \citep{mouschovias83, mouschovias85b}, due to the large lever arm introduced by the fanning out of field lines.
Thus, the analytic studies imply that magnetic braking can have efficiency in the following order: (1) aligned fan-shaped, (2) perpendicular, and (3) aligned uniform configuration. 
Note that the quantitative estimates and equations about magnetic braking are well summarized in \citet{joos12} and \citet{tsukamoto18}.

Here, we define $\theta_0$ as the angle between the rotation axis and magnetic field of the initial cloud. 
For simplicity, we only consider a uniform magnetic field as the initial configuration of prestellar cloud cores.
In analytical studies, two angles $\theta_0 = 0^\circ$ and $90^\circ$ were only considered. 
Figure~\ref{f1} displays such configurations of magnetic field.
We call $\theta_0 = 0^\circ$ the aligned case (panels a--c), and $\theta_0 = 90^\circ$ the perpendicular case (d--h).
Although we give special attention to the perpendicular case, it is also included in the misaligned case ($\theta_0 \ne 0^\circ$).
In addition, in a collapsing cloud (rather than an initial prestellar cloud), the aligned case is further classified into aligned uniform (Figure~\ref{f1}a) and aligned fan-shaped (Figures~\ref{f1}b and c) cases.

The efficiency of magnetic braking is determined by the moment of inertia of the reservoir of angular momentum.
In the star formation process, a massive infalling envelope (or pseudodisk) brakes a less massive rotationally-supported disk to which it is connected through magnetic field lines. 
Thus, the configuration of the magnetic field significantly affects the efficiency of the magnetic braking, because the moment of inertia is determined by the volume (or mass and size) swept by the Alfv\'en waves. 
For example, even in the aligned cases, the moment of inertia for the aligned fan-shaped configuration is larger than that for the aligned uniform configuration because the fan-shaped magnetic field lines connect to the large volume of the infalling envelope that has a large lever arm and a large amount of mass (Figures~\ref{f1}b and c).

The efficiency of magnetic braking also depends on the magnetic field strength, gas density and pitch angle (for the perpendicular case) and opening angle (for the aligned fan-shaped case) of magnetic fields, which would change with time in the star formation process.
For example, in the aligned case, the magnetic field configuration is approximately represented by the aligned uniform configuration in the very early accretion phase, while it gradually transforms to the aligned fan-shaped configuration as the protostar and disk evolve, as described in Figures~\ref{f1}(a)--(c). 
Therefore, it is very difficult using only analytic studies to clearly state which case or configuration dominates during the evolution.
3D simulations are necessary to quantitatively investigate the relation between disk formation and magnetic braking.

In Table~\ref{t1}, we summarize 3D MHD simulations that investigated the misalignment nature in collapsing clouds, in addition to our present study. 
We further comment on each study here.

\begin{deluxetable*}{l|ccccclcc}[t]
\tablecaption{Summary of past studies and this study}
\setlength{\tabcolsep}{2pt}
\tablecolumns{9}
\tablenum{1}
\tablewidth{0pt}
\tablehead{
  \colhead{Authors} &
  \colhead{I/N MHD} &
  \colhead{EOS} &
  \colhead{$\mu_0$} &
  \colhead{$\theta_0$ ($^\circ$)} &
  \colhead{$\alpha_0$} & 
  \colhead{$\beta_0$} &
  \colhead{Sink/EOS} &
  \colhead{Disk}
}
\startdata
\citet{matsumoto04}  &     Ideal & Barotropic & 2, 4, 10      & 0, 45, 70, 80, 90         & 0.5           & 0.02  & Stiff EOS      &     \\
\citet{hennebelle09} &     Ideal & Barotropic & 2, 5, 16      & 0, 10, 20, 60, 90         & 0.25          & 0.03  & Stiff EOS      & Y/N \\ 
\citet{joos12}       &     Ideal & Barotropic & 2, 3, 5, 17   & 0, 20, 45, 70, 80, 90     & 0.25          & 0.03  & Stiff EOS      & Y/N \\
\citet{li13}         &     Ideal & Barotropic & 2.9, 4.9, 9.7 & 0, 45, 90                 & 0.75          & 0.025 & Sink (7\,au)   & Y/N \\
\citet{lewis15}      &     Ideal & Barotropic & 5             & 0, 10, 20, 45, 60, 90     & 0.37          & 0.005 & Sink (1\,au)   &     \\
\citet{masson16}     & Non-Ideal & Barotropic & 2, 5          & 0, 40                     & 0.25          & 0.02  & Stiff EOS      & Y   \\
\citet{tsukamoto18}  & Non-Ideal & RMHD       & 4             & 0, 45, 90                 & 0.2, 0.4, 0.6 & 0.03  & N              & Y   \\
\citet{machida20}    & Non-Ideal & Barotropic & 1.2           & 0, 5, 10, 30, 45, 60, 80, 85, 90 & 0.39 & 0.026  & Stiff EOS & Y   \\
This study           & Non-Ideal & Barotropic & 3             & 0, 10, 30, 45, 60, 80, 90 & 0.2, 0.4, 0.6 & 0.02  & Sink (0.5\,au) & Y   \\
\enddata
\tablecomments{
Column 1: author name(s) for each study;
Column 2: whether non-ideal MHD effects are included (Non-Ideal) or not (Ideal);
Column 3: whether the barotropic equation of state is used (Barotropic) or not (RMHD);
Column 4: mass-to-flux ratio normalized by the critical value used;
Column 5: initial angle between the rotation axis and magnetic field direction;
Columns 6 and 7: ratio of the thermal ($\alpha_0$) and rotational ($\beta_0$) energy relative to the gravitational energy adopted;
Column 8: whether the sink (Sink) or the stiff equation of state (Stiff EOS) is used or not (N), and the sink radius is written when the sink is used;
Column 9: whether the rotationally-supported disk appears (Y) or not, in which Y/N means that the disk appears only when the initial magnetic field is weak.
In the table, a space means no mention about the corresponding item.
}
\label{t1}
\end{deluxetable*}

The ideal MHD approximation was used in all the studies, with the exception of \citet{masson16}, \citet{tsukamoto18} and \citet{machida20}.
Except for \citet{tsukamoto18}, the barotropic equation of state (hereafter EOS) was used to mimic the thermal evolution of star-forming cloud cores, and a sink or stiff EOS was used to accelerate the time evolution in simulations since the time step shortens as the cloud collapse proceeds.
Although \citet{tsukamoto18} calculated the cloud evolution  resolving the protostar without sink particles, they could not calculate the disk evolution for a long duration.

The pioneering work of \citet{matsumoto04} investigated the rate of angular momentum transported by the magnetic braking with a stiff EOS. 
\citet{matsumoto04} calculated the main accretion phase for a short duration and claimed that magnetic braking is more efficient in the perpendicular case than in the aligned case. 
Although they did not comment on whether or not the rotationally-supported disk forms, they showed that the angular momentum in the perpendicular case is less than in the aligned case. 
They calculated the cloud evolution only for $\lesssim\!600$\,yr after the adiabatic core formation, which corresponds to the very early main accretion phase.

Using a stiff EOS, \citet{hennebelle09} calculated the disk evolution for $\sim\!5000$\,yr after the adiabatic core formation and pointed out that the efficiency of magnetic braking decreases as the initial angle difference $\theta_0$ increases. 
In addition, no disk appears when the angle is as small as $\theta_0 \simeq 0^\circ$ even in a weakly magnetized cloud ($\mu_0 = 5$). 
They interpreted that the decrease of magnetic braking efficiency in the cloud with large $\theta_0$ is attributed to the formation of thick pseudodisk. 
When the angle is $\theta_0 \ne 0^\circ$, the twisted magnetic fields amplify the magnetic pressure and thicken the pseudodisk. 
The pseudodisk is a reservoir of the angular momentum and is rotating with a sub-Keplerian velocity.
Thus, the existence of such a thick pseudodisk decreases the efficiency of the magnetic braking \citep[see also][]{ciardi10}.

Using the same numerical code as in \citet{hennebelle09}, \citet{joos12} investigated the disk formation with various parameters of the initial cloud. 
They also concluded that the initial angle difference between $\vect{J}$ and $\vect{B}$ (i.e., $\theta_0 \ne 0^\circ$) promotes the disk formation.
However, unlike \citet{hennebelle09}, they claimed that a fan-like configuration of the magnetic fields, which is realized with a small $\theta_0$, increases the efficiency of the magnetic braking. 
Thus, a large-sized disk tends to appear with a large $\theta_0$.

\citet{li13} also showed that the disk formation is promoted with large angle difference (or large $\theta_0$).
Thus, their result is qualitatively the same as in \citet{hennebelle09} and \citet{joos12}. 
By estimating the magnetic torques, they showed that magnetic braking in the aligned case ($\theta_0 = 0^\circ$) is more efficient than in the misaligned case ($\theta_0 \ne 0^\circ$). 
In addition, they pointed out that the protostellar outflow weakens as $\theta_0$ increases and does not appear in the perpendicular case ($\theta_0 = 90^\circ$). 
Since the outflow directly removes the angular momentum from the central (or disk forming) region, it should greatly affect the formation of a rotationally-supported disk. 
Furthermore, they claimed that magnetic interchange instability \citep{li96,tassis05} possibly disrupts the disk formation. 
However, the sink radius adopted in \citet{li13} is too large to properly investigate the disk formation \citep{machida14,machida16}.

\citet{lewis15} investigated the disk formation in the misaligned cases with a small sink radius of $1$\,au and also showed that the outflow weakens as the angle difference ($\theta_0$) increases. 
However, they did not comment on the formation of rotationally-supported disk in detail.

In these studies \citep{matsumoto04,hennebelle09,joos12,li13,lewis15}, the effects of magnetic braking on the disk formation in both aligned and misaligned cases were investigated in the ideal MHD approximation. 
Then, \citet{masson16} investigated the evolution of both cases with their non-ideal MHD simulations, in which the magnetic field weakens in a high-density region due to ambipolar diffusion. 
Note that they did not include the Ohmic dissipation.
Unlike previous studies, they concluded that misalignment nature does not significantly affect the disk evolution because the magnetic field dissipates in the disk forming region. 
However, they calculated the cloud evolution only with two angle cases ($\theta_0 = 0$ and $40^\circ$).
In addition, the disk radius is larger in the misaligned case ($\theta_0 = 40^\circ$) than in the aligned case ($\theta_0 = 0^\circ$) for a strongly magnetized cloud ($\mu_0 = 2$), while a large-sized disk appears in the aligned case for a weakly magnetized cloud. 
Thus, it is difficult to conclude the effect of misalignment only from this study.

Most recently, \citet{tsukamoto18} pointed out that gravitational stability of the prestellar core significantly affects the results. 
They concluded that the difference between \citet{matsumoto04} and other studies is caused by the difference in $\alpha_0$ that is defined as the ratio of thermal to gravitational energy and controls the mass accretion rate \citep{matsushita17}. 
With a small $\alpha_0$, the cloud is gravitationally very unstable, the mass accretion rate on the disk is high and magnetic field lines are rapidly advected inward in the collapsing cloud, allowing less magnetic braking.
Therefore, in such a case, a large-sized disk tends to appear independent of the parameter $\theta_0$. 
On the other hand, with a large $\alpha_0$, the cloud is marginally gravitationally unstable and the slower collapse allows the magnetic field and magnetic braking to play a significant role in determining the disk size.
However, \citet{tsukamoto18} calculated the disk formation only in the very early accretion phase, because they did not use sink particles \citep[see also][]{machida20}.

To summarize, \citet{matsumoto04} implied that the misalignment nature {\it suppresses} the disk formation, while other studies claimed that it {\it promotes} the disk formation. 
The resolution to this discrepancy is that in addition to the parameter $\theta_0$, the non-ideal MHD effects and the parameter $\alpha_0$ are also significant for considering the disk formation. 
Nevertheless, with non-ideal MHD simulations, no one investigated a long-term evolution of a rotationally-supported disk with investigation of the parameters $\theta_0$ and $\alpha_0$.
The purpose of this study is to further the study of the misalignment problem: whether the misalignment nature promotes or suppresses the disk formation. 
Using non-ideal MHD simulations, we investigate the disk formation using a wide range of parameters as described in Table~\ref{t1}.

\section{Initial Condition and Numerical Settings}
\label{sec:settings}

The initial condition and numerical settings in this study are almost the same as in our past studies \citep{machida06,machida07,machida13}. 
We use our nested grid code, in which the rectangular grids of ($i, j, k$) = ($64$, $64$, $64$) are superimposed \citep{machida04,machida05,machida12}.
The convergence of the numerical resolution is described  in Appendix~\ref{sec:app1}.
We use the index ``$l$'' to describe a grid level. 
The grid size $L(l)$ and cell width $h(l)$ of the $l$-th grid are twice larger than those of ($l+1$)-th grid (e.g., $L(l) = 2L(l+1)$ and $h(l) = 2h(l+1)$).
In the simulation, we resolve the Jeans wave length with at least $16$ cells, and a new finer grid is automatically generated to ensure the Truelove condition \citep{truelove98}.

We solve the resistive MHD equations with the barotropic EOS (equations~1--7 of \citealt{machida12}).
The diffusion rate of Ohmic resistivity is described in \citet{machida07} and \citet{machida12}. 
The differences in numerical settings between past studies and our study are also described in Table~\ref{t1}.

As described in \S\ref{sec:past}, the past studies pointed out the importance of the parameters $\theta_0$ and $\alpha_0$.
We executed two types of simulations to examine the parameter dependence on $\theta_0$ and $\alpha_0$, respectively:
\begin{enumerate}
\item simulations with different angles $\theta_0$ (hereafter we referred to as ``Simulations A'')
\item simulations with different gravitational stabilities $\alpha_0$ (hereafter we referred to as ``Simulations B'').
\end{enumerate}
We executed $16$ simulations in total. 
Each simulation required $2$--$4$ months of wall-clock-time. 
Since the simulations take a very long time to run, and were performed over the course of more than one year, all simulations were not coordinated to have exactly the same initial conditions.
Therefore, Simulations A and B differ slightly even when the parameters $\theta_0$ and $\alpha_0$ are the same (for details, see below). 
For easy understanding, the initial conditions of Simulations A and B are separately described in the following subsections (\S\ref{sec:simA} and \ref{sec:SimB}).

\begin{deluxetable*}{c|rcclcccrrcc}[t]
\tablecaption{Parameters and cloud physical quantities of initial clouds and calculation results}
\tablecolumns{12}
\tablenum{2}
\tablewidth{0pt}
\tablehead{
    \colhead{Model} &
    \colhead{$\theta_0$} &
    \colhead{$\alpha_0$} &
    \colhead{$f$} &
    \colhead{$M_{\rm cl}$} &
    \colhead{$R_{\rm cl}$} & 
    \colhead{$B_0$} &
    \colhead{$\Omega_0$} &
    \colhead{$R_{\rm disk}$} &
    \colhead{$R_{90\%}$} &
    \colhead{$M_{\rm disk}$} &
    \colhead{$M_{\rm sink}$} \\
    \colhead{} &
    \colhead{($^\circ$)} &
    \colhead{} &
    \colhead{} &
    \colhead{($\msun$)} &
    \colhead{(au)} & 
    \colhead{($10^{-5}$\,G)} &
    \colhead{($10^{-13}\,{\rm s}^{-1}$ )} &
    \colhead{(au)} &
    \colhead{(au)} & 
    \colhead{($\msun$)} &
    \colhead{($\msun$)}
}
\startdata
A1 &  0 &     &     &      &                 &    &     &  10.5 &  7.9 & 0.019 & 0.098 \\
A2 & 10 &     &     &      &                 &    &     &  28.8 & 13.8 & 0.044 & 0.089 \\
A3 & 30 &     &     &      &                 &    &     &  36.3 & 21.9 & 0.050 & 0.104 \\
A4 & 45 & 0.4 & 2.0 & 1.0  & $5.9\times10^3$ & 43 & 1.5 &  50.1 & 22.9 & 0.060 & 0.100 \\
A5 & 60 &     &     &      &                 &    &     &  91.2 & 33.1 & 0.051 & 0.116 \\
A6 & 80 &     &     &      &                 &    &     &  69.2 & 22.9 & 0.048 & 0.124 \\
A7 & 90 &     &     &      &                 &    &     &  57.5 & 30.2 & 0.032 & 0.136 \\
\hline 
B1 &  0 &     &     &      &                 &    &     &  27.5 & 15.1 & 0.090 & 0.222 \\
B2 & 45 & 0.2 & 4.2 & 2.5  & $6.2\times10^3$ & 97 & 2.1 &  75.8 & 26.3 & 0.152 & 0.278 \\
B3 & 90 &     &     &      &                 &    &     & 109.6 & 28.8 & 0.141 & 0.296 \\
\hline 
B4 &  0 &     &     &      &                 &    &     &  12.0 &  9.1 & 0.025 & 0.110 \\
B5 & 45 & 0.4 & 2.1 & 1.2  & $6.2\times10^3$ & 48 & 1.5 &  57.5 & 26.3 & 0.068 & 0.127 \\
B6 & 90 &     &     &      &                 &    &     &  72.4 & 20.9 & 0.049 & 0.151 \\
\hline 
B7 &  0 &     &     &      &                 &    &     &   9.5 &  6.9 & 0.011 & 0.084 \\
B8 & 45 & 0.6 & 1.4 & 0.82 & $6.2\times10^3$ & 26 & 1.2 &  47.8 & 20.9 & 0.044 & 0.089 \\
B9 & 90 &     &     &      &                 &    &     &  60.3 & 21.9 & 0.016 & 0.117 \\
\enddata
\tablecomments{
Column 1: model name;
Columns 2 and 3: parameter $\theta_0$ (the initial angle difference between the rotation axis and magnetic field) and $\alpha_0$ (the ratio of thermal to gravitational energy);
Column 4: density enhancement factor;
Columns 5 and 6: initial cloud mass and radius;
Columns 7 and 8: initial magnetic field strength and angular rotation rate;
Columns 9--12: disk radius, radius of $M_{\rm enc} = 0.9 M_{\rm disk}$ (for definitions, see \S 4), disk mass, and protostellar mass at the end of the simulation ($t_{\rm ps} = 5000$\,yr).
}
\label{t2}
\end{deluxetable*}

In both Simulations A and B, we adopted a Bonnor-Ebert (BE) density profile for the initial cloud. 
A rigid rotation is adopted within the BE sphere, and a uniform magnetic field is imposed in the whole computational domain.
In Cartesian coordinates, the direction of the global magnetic field is always parallel to the $z$-axis, while the rotation direction is inclined from the $z$-axis to the $x$-axis by an angle of $\theta_0$.

The spatial resolution is the same among all simulations.
The grid size and cell with of the $l = 1$ grid are $L(1) = 1.9\times10^5$\,au and $h(1) = 2.9\times10^3$\,au, respectively.
We set the maximum grid level as $l = 14$ and the finest grid has $L(14) = 24$\,au and $h(14) = 0.36$\,au, respectively.

To accelerate the time evolution, we adopt a sink cell technique \citep[for details, see][]{machida10,machida13}.
We set a threshold density $n_{\rm thr}$ and sink accretion radius $r_{\rm sink}$. 
In the collapsing cloud, we define the protostar formation epoch to begin when the gas density exceeds $n_{\rm thr}$. 
After protostar formation or sink creation, we remove the gas exceeding $n_{\rm thr}$ in the region of $r < r_{\rm sink}$ and add its mass to the gravitational potential of the protostar \citep{machida10}. 
In both Simulations A and B, we set $n_{\rm thr} = 10^{14}\,\cc$ and $r_{\rm sink} = 0.5$\,au, respectively.
In the following, we explain details for each initial prestellar cloud in Simulations A and B.
These parameters are summarized in Table~\ref{t2}.

\subsection{Simulations A}
\label{sec:simA}

We prepare the BE density profile $\rho_{\rm BE}(r)$ with the central density $n_{\rm c,0} = 5 \times 10^5\,\cc$ and temperature $T_0 = 10$\,K.
To promote the cloud contraction, we enhance the density by $f$ (density enhancement factor) as $\rho(r) = f \rho_{\rm BE}(r)$ and adopted it as the density profile of the initial cloud.
With $f = 2$, the mass and size of the initial cloud are $M_ {\rm cl} = 1\,\msun$ and $R_{\rm cl} = 5.9\times10^3$\,au, respectively.
The magnetic field strength is adjusted to have a mass-to-flux ratio $\mu_0 = 3$, where $\mu_0$ is normalized by the critical value $(4 \pi^2 G)^{-1/2}$.\footnote{The mass-to-flux ratio becomes $\mu_0 = 3.797$ in the definition of \cite{mouschovias76}, in which the normalized factor $\sim\!(6.408 \pi^2 G)^{1/2}$ is adopted.}
The strength of the uniform magnetic field is $B_0 = 4.3\times10^{-5}$\,G.
A rigid rotation of $\Omega_0 = 1.5\times10^{-13}\,{\rm s}^{-1}$ is adopted.
With these settings, the ratio of thermal ($\alpha_0$) and rotational ($\beta_0$) energies to the gravitational energy of the initial cloud are $\alpha_0 = 0.4$ and $\beta_0 = 0.02$, respectively.
For the parameter $\theta_0$, we adopt $\theta_0 = 0$, $10$, $30$, $45$, $60$, $80$, and $90^\circ$.
Thus, we executed seven simulations in total for Simulations A.

\subsection{Simulations B}
\label{sec:SimB}

Since $\alpha_0 \propto c_{s,0}^2\,M_{\rm cl}^{-1}$ (where $c_{s,0}$ is the sound speed and constant in the initial cloud), we can vary the value of parameter $\alpha_0$ by changing the initial cloud density (or initial cloud mass $M_{\rm cl}$).
The initial clouds prepared for Simulations B are almost the same as those for Simulations A but varying the density enhancement factor $f$ in order to change the parameter $\alpha_0$.
All the models in Simulations B have the same cloud radius $R_{\rm cl} = 6.2\times10^3$\,au, which is slightly larger than in Simulations A.
The mass differs in each model as listed in Table~\ref{t2}, as we changed the initial cloud density.
The initial angular velocity $\Omega_0$ and uniform magnetic field $B_0$ are adjusted to yield $\beta_0 = 0.02$ and $\mu_0 = 3$, respectively.
Thus, models with different $\alpha_0$ have the same non-dimensional parameters $\beta_0$ and $\mu_0$ but different dimensional parameters $\Omega_0$ and $B_0$.

\section{Results}
\label{sec:results}

As described in \S\ref{sec:settings}, we executed two types of simulations.
Simulations A only changed the parameter $\theta_0$ to focus on the effect of the angle difference between rotation axis and global field on the disk formation. 
Simulations B changed both the parameters $\theta_0$ and $\alpha_0$ in order to investigate the effect of the initial gravitational stability of the cloud.

\subsection{Simulations A: $\theta_0$ dependence}
\label{sec:SimAresults}

\begin{figure}[t]
  \begin{center}
    \includegraphics[width=1.0\columnwidth]{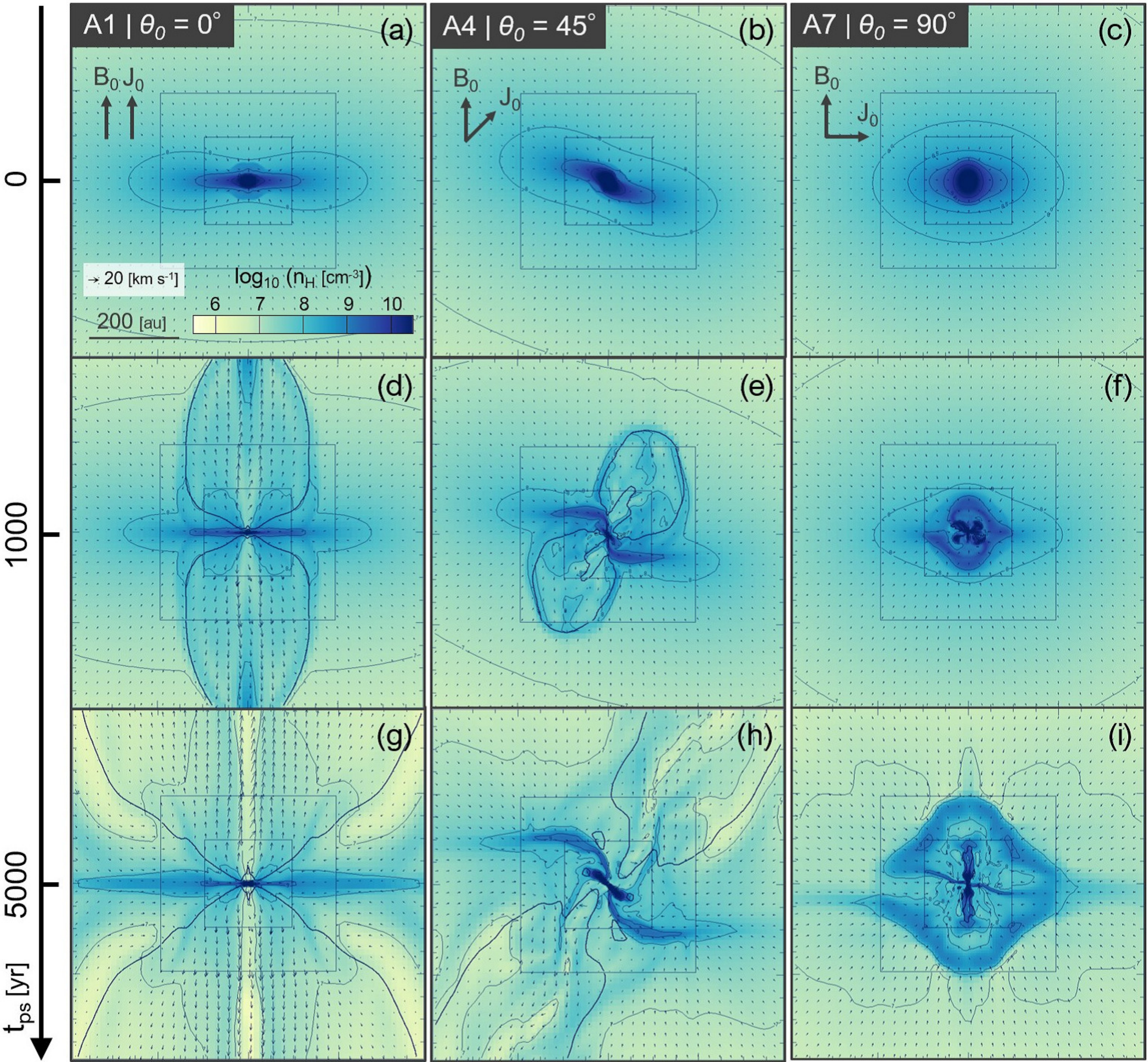}
  \end{center}
\caption{
Density (color and contours) and velocity (arrows) distributions on the $y = 0$ plane for models A1 ($\theta_0 = 0^\circ$; panels a, d, and g), A4 ($\theta_0 = 45^\circ$; b, e, and h), and A7 ($\theta_0 = 90^\circ$; c, f, and i) at $t_{\rm ps} = 0$, $1000$, and $5000$\,yr after protostar formation.
The box size is $780$\,au.
The squares in each panel represent a boundary of a nested grid.
}
\label{f2}
\end{figure}

\begin{figure}[t]
  \begin{center}
    \includegraphics[width=1.0\columnwidth]{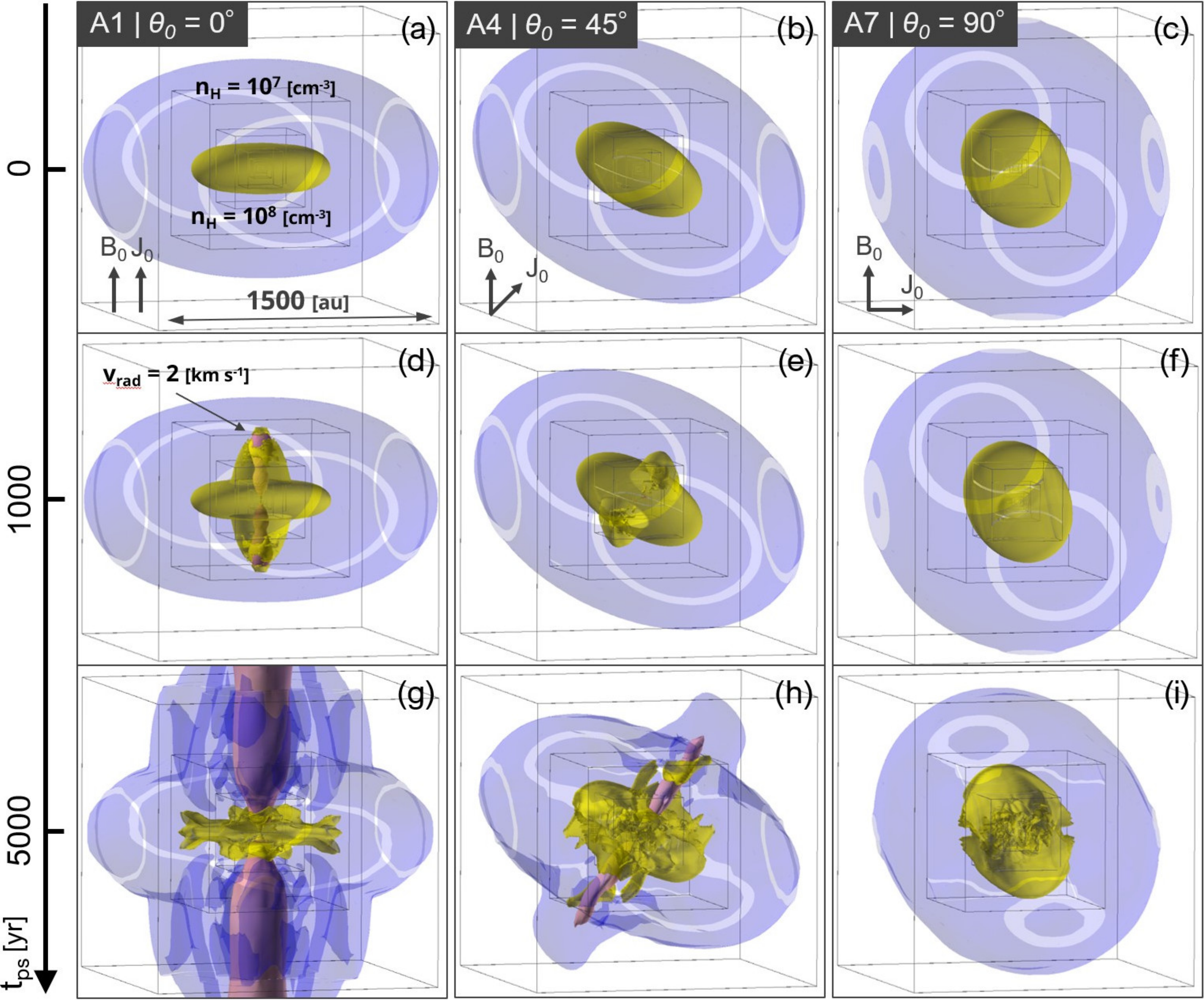}
  \end{center}
\caption{
Three-dimensional structures for the same models as in Figure~\ref{f2}.
We depict two isodensity contours of $\nh = 10^7\,\cc$ (purple) and $10^8\,\cc$ (yellow) and one isovelocity contour of radial velocity $v_{\rm rad} = 2\,\kms$ (red).
The box size is $780$\,au.
The squares in each panel denote a boundary of a nested grid.
The white ellipse in each panel corresponds to the cutting plane of the boundary of the outermost grid surface. 
}
\label{f3}
\end{figure}

\begin{figure*}[t]
  \begin{center}
    \includegraphics[width=0.7\linewidth]{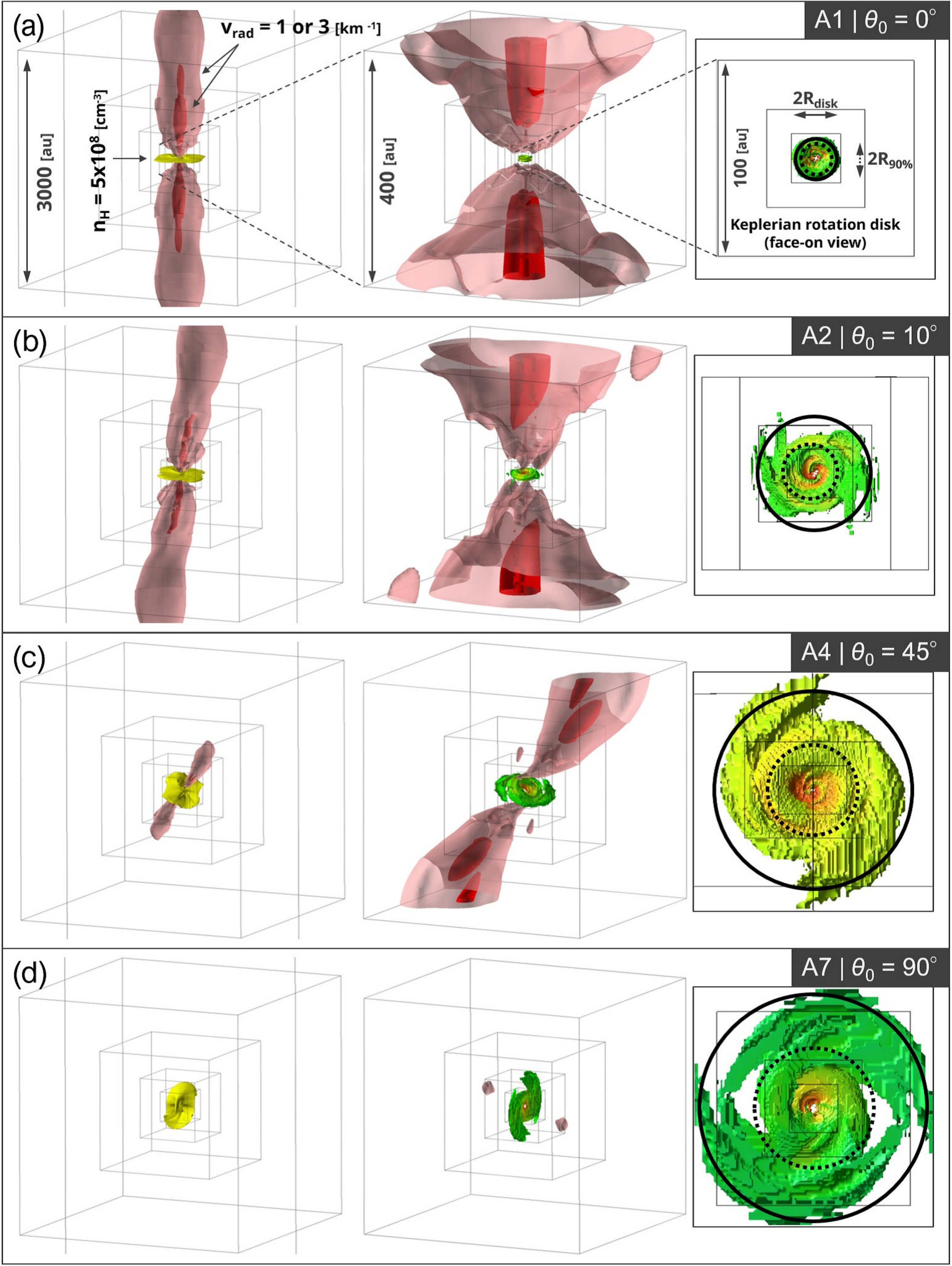}
  \end{center}
\caption{
Three-dimensional view of the outflow, pseudodisk, and rotationally-supported disk with a box size of $3000$\,au (left column), $400$\,au (middle), and $100$\,au (right) for models A1 ($\theta_0 = 0^\circ$; panel a), A2 ($\theta_0 = 10^\circ$; b), A4 ($\theta_0 = 45^\circ$; c), and A7 ($\theta_0 = 90^\circ$; d) at $t_{\rm ps} = 5000$\,yr.
The outflow is represented by the isovelocity surfaces of radial velocity $v_{\rm rad} = 1$ (pink) and $3\,\kms$ (red), respectively. 
In the left panels, the pseudodisk is represented by the yellow surface corresponding to an isodensity of $n_{\rm H} = 5 \times 10^8\,\cc$.
In the middle and right panels, the rotationally-supported disk is plotted by a colored surface (the color represents the density on the disk surface). 
In each right panel, the viewing angle is adjusted to be face-on to the disk surface.
The disk radii $R_{\rm disk}$ and $R_{90\%}$ are indicated by the solid and dashed circles, respectively.
The nested squares in each panel indicate the boundaries of a nested grid.
}
\label{f4}
\end{figure*}

Figure~\ref{f2} shows the time evolution for three models at $t_{\rm ps} = 0$, $1000$, and $5000$\,yr, in which $t_{\rm ps} = 0$ at the sink creation epoch corresponding to the protostar formation.
In the alignment model A1 ($\theta_0 = 0^\circ$; the left column of Figure~\ref{f2}), the outflow gradually evolves and has a size of $\sim\!1000$\,au at $t_{\rm ps} = 5000$\,yr.
Panels (d) and (g) show an oblate structure with a density of $\sim\!10^8\,\cc$ corresponds to the pseudodisk. 
We confirm that the pseudodisk becomes thin with time. 
The high-density region ($\gtrsim\!10^9\,\cc$) corresponds to the rotationally-supported (or dense) disk.\footnote{
Below we properly define the rotationally-supported disk with Figure~\ref{f4}. Before that time, we call the disk ``the high-density disk'' or simply ``the disk''. Actually, the high-density disk does roughly correspond to the rotationally-supported (or Keplerian) disk.}
The high-density disk becomes large with time.
The bipolar outflow and geometrically thin and dense disk are clearly reproduced in the alignment model.

Next we focus on the misalignment model A4 ($\theta_0 = 45^\circ$; the middle column of Figure~\ref{f2}).
The outflow and disk system also appear in the misalignment model as seen in panels (e) and (h).
However, the outflow direction is not aligned with the global (or initial) field direction that is parallel to the $z$-direction.
The outflow direction is almost perpendicular to the disk direction, which indicates that the outflow propagates along the disk normal direction roughly corresponding to the initial rotation direction ($\vect{J_0}$).
Thus the outflow propagation direction is not always coincident with the global $B$-field direction ($\vect{B_0}$).
Misalignments between the inferred magnetic field direction and protostellar outflows have been observed by \cite{hull13}.
In our simulations, the misalignment is clearly produced in the models with $\theta_0 \ne 0^\circ$, in which the global field is not aligned with the initial rotation axis of the prestellar cloud core.

Figure~\ref{f2} also indicates that the outflow in the misalignment model (middle column) is weaker than in the alignment model (left column). 
The effect of $\theta_0$ on the outflow strength is discussed below.
In addition, the misalignment model shows a complex structure of the pseudodisk. 
Figures~\ref{f2}(e) and (h) show a spiral structure that is formed from the pseudodisk twisted by rotation.
The inner dense disk is enclosed by a twisted pseudodisk (Figure~\ref{f2}h).

Finally, we focus on the perpendicular model A7 ($\theta_0 = 90^\circ$; the right column of Figure~\ref{f2}).\footnote{The mirror symmetries in both the $x$- and $z$-directions are somewhat broken in Figures~\ref{f2}(f) and (i) but do not qualitatively affect the analyzed properties of the disk and outflow (see Appendix~\ref{sec:app2}).}
In this model, no strong outflow appears until the end of the calculation\footnote{A very weak outflow does appear in the perpendicular model as shown in Figure~\ref{f9}.}, while a large-sized dense disk, corresponding to the rotationally-supported disk (for details, see below), does appear. 
The normal direction of the dense disk corresponds to the rotation axis of the initial cloud core.
In addition, the inner dense disk is wrapped by a relatively low-density spiral that corresponds to the pseudodisk twisted strongly by the rotation.

Figure~\ref{f3} shows a time-sequence of structures in three dimensions for the same models.
The figure indicates that the outflow (red isovelocity surface) and dense disk (yellow isodensity surface) system is enclosed by a less dense pseudodisk (purple isodensity surface) in the alignment (left column) and misalignment models (middle), while only a dense disk (yellow) is embedded in a nearly spherical structure (purple) in the perpendicular model (right). 
In addition, we can confirm that the pseudodisk (purple) is more distorted in the alignment models than in the perpendicular model.
In other words, the misalignment models with $\theta_0 = 45$ and $90^\circ$ have a nearly spherical structure (or a nearly spherical pseudodisk) around a geometrically thin and dense disk.
This was also pointed out by \cite{hennebelle09}.

\begin{figure*}[t]
  \begin{center}
    \includegraphics[width=0.9\linewidth]{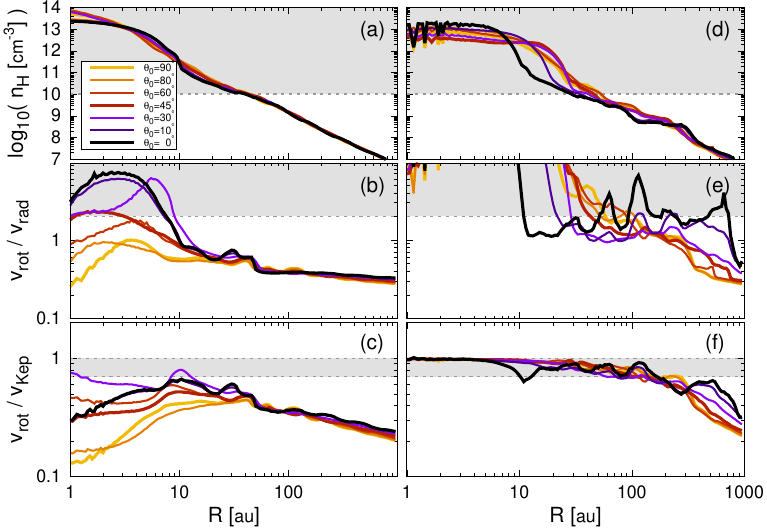}
  \end{center}
\caption{
Radial profiles for all models in Simulations A at $t_{\rm ps} = 0$\,yr (left panels) and $5000$\,yr (right), respectively: (panels a and d) number density, (b and e) ratio of the rotational to the radial velocity, and (c and f) ratio of the rotational to the Keplerian velocity.
The gray zone in each panel corresponds to each criteria (i)--(iii) for identifying the rotationally-supported disk described in the text.
}
\label{f5}
\end{figure*}

To compare the spatial scales between models with different $\theta_0$, Figure~\ref{f4} plots 3D views of the outflow, pseudodisk, and rotationally-supported disk at $t_{\rm ps} = 5000$\,yr for models with $\theta_0 = 0$, $10$, $45$, and $90^\circ$. 
The figure indicates that, with the increment of $\theta_0$, the outflow gradually weakens while the disk seems to increase its size. 
However, it is difficult to distinguish the rotationally-supported (or Keplerian) disk from the pseudodisk only with the density distribution, because the pseudodisk also has a disk-like structure in the high density region. 
To clearly identify the rotationally-supported disk in the computational domain, we imposed the following criteria:
\renewcommand{\labelenumi}{(\roman{enumi})}
\begin{enumerate}
\item the number density is larger than $n>10^{10}\,\cc$,
\item the rotation velocity is greater than the radial velocity $\vert v_\phi \vert > f_{\rm rot} \vert v_{\rm rad} \vert$ by a factor $f_{\rm rot}$, where $f_{\rm rot} = 2$ is adopted, 
\item the rotation velocity is in the range of $0.7 \le \vert v_{\phi} / v_{\rm kep} \vert \le 1.0$, where $v_{\rm kep} = [G (M_{\rm ps} + M_{\rm enc}(r))/r]^{1/2}$ is the Keplerian velocity. 
\end{enumerate} 
Using the sink mass $M_{\rm sink}$\footnote{
At every timestep, we estimated the gas falling onto the sink cell and added it to $M_{\rm sink}$. 
}, which is equated to the protostellar mass $M_{\rm ps}$, and the enclosed mass $M_{\rm enc}(r)$ within a radius $r$, we estimate the Keplerian velocity $v_{\rm kep}$.
The procedure to identify the rotationally-supported disk is almost the same as that used in \citet{joos12}.
Note that although we imposed a lower limit of Keplerian velocity $0.7 v_{\rm kep}$ in criterion (iii), the disk physical quantities such as mass and size do not significantly depend on the lower limit.
The disk-like structure delineated by the green color in the middle and right panels of Figure~\ref{f4} corresponds to the region where the criteria (i)--(iii) are satisfied.
Thus, it corresponds to the rotationally-supported (or Keplerian) disk.

\begin{figure*}[t]
  \begin{center}
    \includegraphics[width=0.9\linewidth]{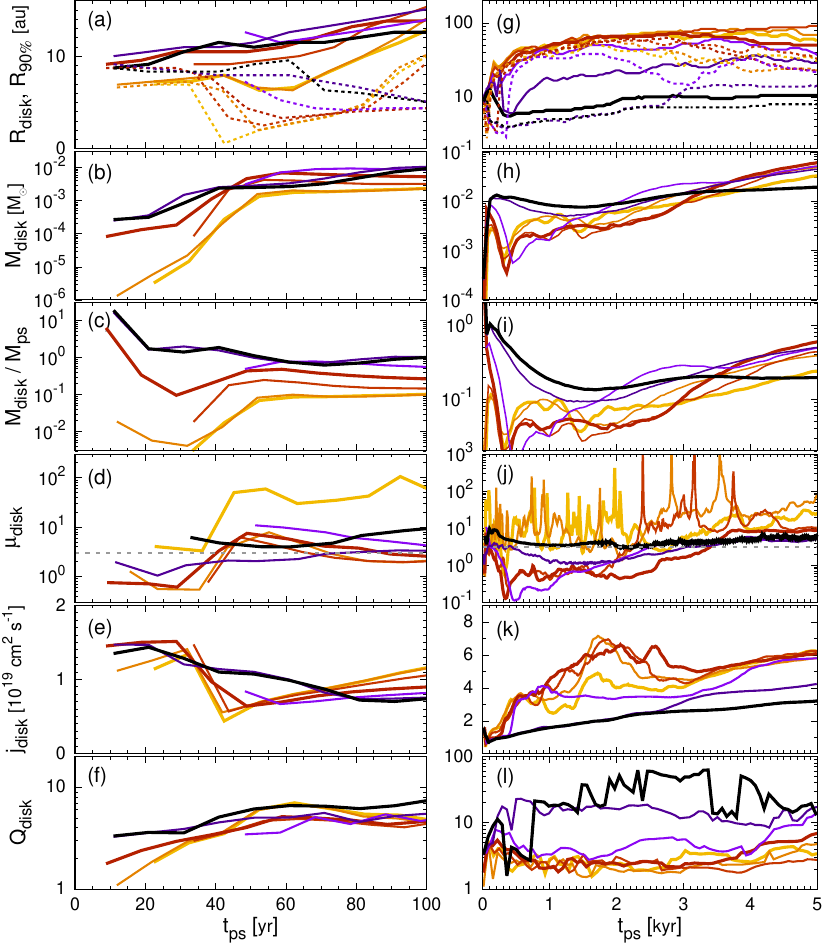}
  \end{center}
\caption{
Time evolution of disk properties for all models in Simulations A during the first $100$\,yr (left panels) and $5000$\,yr (right) after the protostar formation, respectively: (panels a and g) radius of $M=M_{\rm tot}$ ($R_{\rm disk}$, solid lines) and $M=0.9M_{\rm tot}$ ($R_{90\%}$, dotted lines), (b and h) disk mass, (c and i) ratio of disk mass to protostellar mass, (d and j) magnetization (or normalized mass-to-flux ratio), (e and k) specific angular momentum, and (f and l) Toomre Q parameter averaged over the disk.
The dashed line in panel (d) plots the initial mass-to-flux ratio, $\mu_0 = 3$.
}
\label{f6}
\end{figure*}

The right panels of Figure~\ref{f4} show that the rotationally-supported disks have a spiral or complex structure near the disk outer edge. 
The spiral structure occupies a large extent especially in the misalignment models. 
The disk outer edge shows a high time variability and its structure changes in a short duration. 
Thus, a subtle structural change near the disk outer edge induces a significant change in the disk size. 
To exclude apparent or transient size changes of the disk, we defined the mass-weighted disk size, $R_{\rm disk}$, inside which the rotationally-supported-disk mass is contained. 
The radius of the mass-weighted disk is derived through the following procedure:
\begin{enumerate}
\renewcommand{\labelenumi}{\arabic{enumi})}
\item 
The total mass of the rotationally-supported disk, $M_{\rm tot}$, is derived by  integrating over the entire disk,\footnote{The rotationally-supported disk is identified with the criteria (i)--(iii).} 
\item 
then, the mass of the rotationally-supported disk is radially integrated from the center until the integrated mass $M_{\rm disk}$ reaches $M_{\rm tot}$, and, 
\item 
finally, the disk radius $R_{\rm disk}$ is determined as the radius inside which $M_{\rm tot}$ is contained.
\end{enumerate}
The disk radius $R_{\rm disk}$ is represented by the solid circle in the right panels of Figure~\ref{f4}. 
From these panels, we can confirm that the spiral or complex structure is contained within $R_{\rm disk}$.
We use $R_{\rm disk}$ as a typical radius of the rotationally-supported disk and simply call it the disk radius. 
In addition, we define another radius $R_{90\%}$ inside which $90\%$ of the disk mass is contained. It is represented by the dashed circle in Figure~\ref{f4}.
The size difference between $R_{\rm disk}$ and $R_{90\%}$ means that a large part of the disk mass is concentrated in the central region of the disk.

Figure~\ref{f5} plots the radial profiles of number density, ratio of rotational to radial velocity, and ratio of rotational to Keplerian velocity for models A1--A7 at $t_{\rm ps} = 0$\,yr (the protostar formation epoch) and $5000$\,yr (the end of the simulation), in which each value is azimuthally averaged. 
At the protostar formation epoch, the models have almost the same density distribution (Figure~\ref{f5}a).
In addition, Figure~\ref{f5}(b) means that for all the models, the rotational velocity dominates or is comparable to the radial velocity even though the rotational velocity does not yet reach the Keplerian velocity.
At the protostar formation epoch, all the models satisfy the disk identification criteria (i) $n > 10^{10}\,\cc$ (Figure~\ref{f5}a) and some models satisfy the criteria (ii) $\vert v_{\phi} \vert > 2 \vert v_{\rm rad} \vert$ (Figure~\ref{f5}b).
However, no model satisfies the third criteria (iii) $0.7 \le \vert v_{\phi} / v_{\rm Kep} \vert \le 1.0$ (Figure~\ref{f5}c).
Therefore, no Keplerian disk forms at this epoch.

On the other hand, at $t_{\rm ps} = 5000$\,yr, all the models satisfy the identification criteria for a rotationally-supported disk (i)--(iii), as seen in Figures~\ref{f5}(d--f). 
Especially, Figure~\ref{f5}(f) indicates that the rotation velocities are in rough agreement with the Keplerian velocity at $R \lesssim 10$--$100$\,au. 
Hence, in all models, the Keplerian disk appears {\it after} protostar formation.

Next, we show the disk properties for all models in Simulations A. 
Top three rows in Figure~\ref{f6} shows the disk radii $R_{\rm disk}$ and  $R_{90\%}$, disk mass $M_{\rm disk}$, and ratio of disk mass to protostellar mass $M_{\rm disk} / M_{\rm ps}$ against the elapsed time after protostar formation ($t_{\rm ps}$).
The time evolution for the first $100$\,yr after protostar formation are plotted in the left panel, while those for 5000\,yr (by the end of the simulations) are plotted in the right panel. 
In previous studies, \citet{matsumoto04} showed that the disk is larger in the aligned case ($\theta_0 = 0^\circ$) than in the perpendicular case ($\theta_0 = 90^\circ$).
On the other hand, \citet{joos12} showed the opposite result; the disk is larger in the misaligned case ($\theta_0 \ne 0^\circ$) than in the aligned case. 
The disk evolution for the first $\sim\!600$\,yr was calculated in \citet{matsumoto04}, while up to $\sim\! 5000$\,yr were investigated by \citet{joos12}. 
The parameters of initial clouds also differ between two studies.
Here we investigate both a short and long timescale of the disk evolution, so we prepared two sets of panels in Figure~\ref{f6}.

During the very early phase after protostar formation ($t_{\rm ps} < 100$\,yr), the disk tends to have a small radius and small mass in models with a large $\theta_0$ compared to models with a small $\theta_0$.
The disk radius $R_{\rm disk}$ for the model with $\theta_0 = 90^\circ$ is the smallest among the models for $t_{\rm ps} \lesssim 100$\,yr, while it becomes the largest after that (Figure~\ref{f6}a). 
The radii of $M = 0.9\,M_{\rm tot}$, $R_{90\%}$, are smaller than $R_{\rm disk}$ but the dependence of $R_{90\%}$ evolution on $\theta_0$ is similar to $R_{\rm disk}$.
The disk mass for the model with $\theta_0 = 90^\circ$ is also the smallest for $t_{\rm ps} \lesssim 100$\,yr (Figure~\ref{f6}b). 
In addition, Figure~\ref{f6}(c) indicates that the disk mass for the model with $\theta_0 \ge 80^\circ$ is one or two orders of magnitude less than the protostellar mass, while the disk mass for models with $\theta_0 \le 30^\circ$ is comparable to the protostellar mass. 
Thus, during the early phase, the disk radius and mass for the alignment model are larger than that for the perpendicular model, which agrees well with the results of \citet{matsumoto04}.

Both the radius- and mass-magnitude relations show a variation with time. 
Figure~\ref{f6}(g) shows that the disk radius for the alignment model is as small as $r_{\rm disk} \sim 10$\,au by the end of the simulation, which agrees with \citet{machida14} and \citet{machida19}. 
On the other hand, during $t_{\rm ps} \lesssim 5000$\,yr, the disk radius increases to $10$--$100$\,au in the misalignment models.
Thus, the radius-magnitude relation is reversed at these times. 
In Figure~\ref{f6}(g), the dependence of disk radius on $\theta_0$ is also the same for $R_{\rm disk}$ and $R_{90\%}$ (see also the final values in Table~\ref{t2}).
Figures~\ref{f6}(h) shows that the magnitude relation of the disk mass is also reversed at $t_{\rm ps} \sim 3000$\,yr, as the disk mass for the perpendicular model becomes the largest and that for the alignment model becomes the smallest among the models.
This tendency is in good agreement with \citet{joos12} and \citet{hennebelle09}, who claimed that the radius and mass are maximized for the perpendicular case.

Figures~\ref{f6}(d) and (j) show the time evolution of the magnetization (normalized mass-to-flux ratio) in the disk, $\mu_{\rm disk}$.
The magnetization in the disk is determined by the balance between the  amplification and dissipation of the magnetic field.
The magnetic field in the disk is amplified by the cloud contraction and the accretion of magnetized matter, while it dissipates by Ohmic dissipation.
In the early phase, $\mu_{\rm disk}$ is lower in misaligned models.
Note that the mass-to-flux ratio of the perpendicular model ($\theta_0=90^\circ$) is largest among models, which may be attributed to the geometry of the accretion in the early accretion phase \citep{tsukamoto18}.
In the later phase, $\mu_{\rm disk}$ is lower with decreasing $\theta_0$.
Although the time evolution of $\mu_{\rm disk}$ is not very simple, the disk tends to have a weak magnetic field when the disk is large or massive. 
This is natural because the dissipation of magnetic field is efficient in the high-density disk region.

\begin{figure}[t]
  \begin{center}
    \includegraphics[width=1.00\columnwidth]{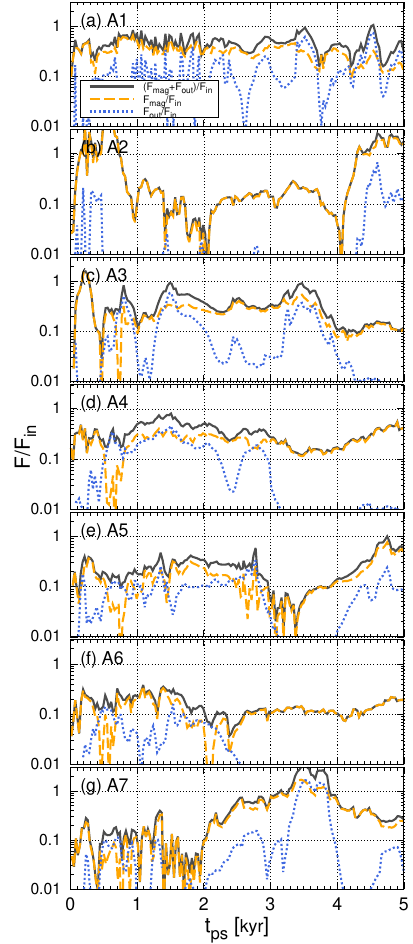}
  \end{center}
\caption{
Time evolution of the outgoing fluxes of the angular momentum  (Equations~\ref{eq:FluxMag} and \ref{eq:FluxOutflow}) normalized by the incoming flux (Equation~\ref{eq:FluxAccretion}) for all models in Simulations A until $5000$\,yr after the protostar formation: $(F_{\rm mag} + F_{\rm out})/F_{\rm in}$ (solid lines), $F_{\rm mag}/F_{\rm in}$ (dashed), and $F_{\rm out}/F_{\rm in}$ (dotted).
}
\label{f7}
\end{figure}

The angular momentum of the disk is carried away by the outflows and magnetic field.
To quantify the relative importance of the mechanisms, we measure the integral fluxes of angular momentum \citep[for details, see Section 5.4.1 in][]{joos12} carried away by the magnetic torque,
\begin{equation}
  F_{\rm mag} = \left| \int_{\rm disk} r \frac{B_{\phi}}{4 \pi} {\bf B} \cdot d{\bf S} \right| \ ,
  \label{eq:FluxMag}
\end{equation}
carried away by the outflows,
\begin{equation}
  F_{\rm out} = \left| \int_{\rm disk} \rho r  v_{\phi} {\bf v} \cdot d{\bf S} \right| \ , \ \text{for} \ {\bf v} \cdot d{\bf S} > 0,
  \label{eq:FluxOutflow}
\end{equation}
and carried in by the accretion flow,
\begin{equation}
  F_{\rm in} = \left| \int_{\rm disk} \rho r v_{\phi} {\bf v} \cdot d{\bf S} \right| \ , \ \text{for} \ {\bf v} \cdot d{\bf S} < 0.
  \label{eq:FluxAccretion}
\end{equation}
We calculate the total fluxes over the surface $S$, of a cylinder having a radius corresponding to the Keplerian disk, in which the disk rotation axis is adopted as the axis of the cylinder. 
The height of the cylinder is set to enclose the whole region of the disk. 
Thus, both the axis and size of the cylinder change with time.  
Figure~\ref{f7} plots the outgoing fluxes of the angular momentum normalized by the incoming flux.
Although the angular momentum transferred by the outflow sometimes dominates that by the magnetic torque, the magnetic torque contributes mainly to the angular momentum transfer. 
Thus, the magnetic torque (i.e., magnetic braking) is a more important factor for disk formation, which is consistent with \citet{joos12}.
These incoming and outgoing fluxes should primarily determine the total angular momentum of the disk (Figures~\ref{f6}e and k) and the disk properties.
We tentatively calculated the time evolution of the angular momentum assuming a cylindrical surface, as according to \citet{joos12}.
However, it is very difficult to more precisely estimate the angular momentum transported from the central region where other mechanisms such as gravitational and thermal pressure torques should  play a role for transporting the angular momentum.
Although further detailed analysis of the disk angular momentum (transport) is necessary, it is beyond the scope of this study.

\begin{figure}[t]
  \begin{center}
    \includegraphics[width=0.95\columnwidth]{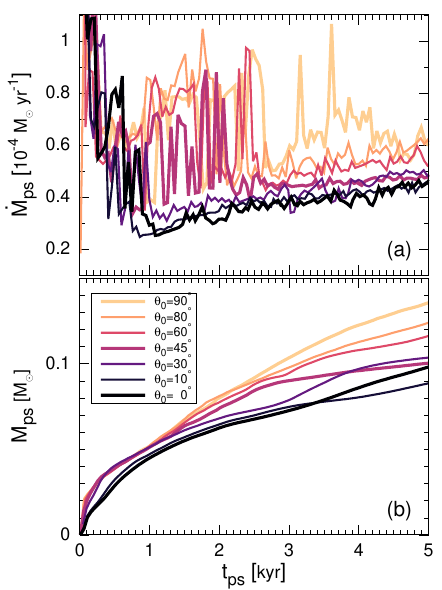}
  \end{center}
\caption{
Mass accretion rate (top panel) and protostellar mass (bottom) versus the elapsed time after protostar formation for all models in Simulations A.
}
\label{f8}
\end{figure}

\begin{figure}[t]
  \begin{center}
    \includegraphics[width=0.95\columnwidth]{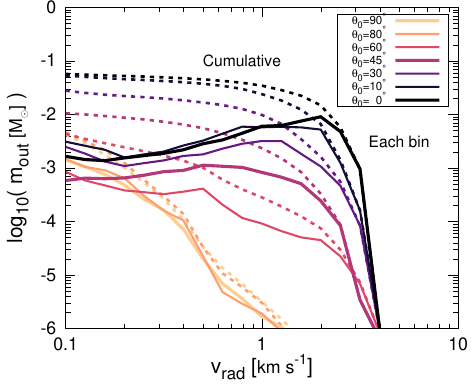}
  \end{center}
\caption{
Mass versus the radial velocity at the end of the simulation ($t_{\rm ps} = 5000$\,yr) for all models in Simulations A.
Solid and broken lines plot the mass in each velocity bin ($m_{\rm out}$; Equation~\ref{eq:1}) and the cumulative mass ($M_{\rm out}$; Equation~\ref{eq:2}), respectively.
}
\label{f9}
\end{figure}

To confirm the disk stability, we estimate the mass-weighted Toomre Q parameter \citep{toomre64} which is averaged over the entire disk $R < R_{\rm disk}$ as
\begin{equation}
  Q_{\rm disk} = \frac{\int_{\rm disk} (c_{\rm s} \kappa)/(\pi G \Sigma) \Sigma \, dS}{\int_{\rm disk} \Sigma \, dS} \ ,
  \label{eq:ToomreQ}
\end{equation}
where $c_s$ is the sound speed, $G$ is the gravitational constant, $\Sigma$ is the surface density, and 
\begin{equation}
  \kappa = 4 \Omega^2 + r\frac{d\Omega^2}{dR} \ 
  \label{eq:EpicyclicFrequency}
\end{equation}
is the epicyclic frequency, in which $\Omega$ is the angular velocity.
Figures~\ref{f6}(f) and (l) plot the time evolution of $Q_{\rm disk}$.
Note that the $Q$ parameter estimated here is an average of the entire Keplerian disk, and not a local quantity.
However, it is plausible to use it to discuss the gravitational instability among the models.
During the first $100$\,yr, $Q_{\rm disk}$ gradually increases, which means that the disk gradually stabilizes.
After $Q_{\rm disk}$ has a local peak around $t_{\rm ps}\simeq100$\,yr (Figure~\ref{f6}l), it decreases with time in the misalignment models ($\theta_0\ne 0^\circ$). 
On the other hand, we can see a strong oscillation in the $Q$ parameter for the aligned model $\theta_0=0^\circ$, as seen in \citet{tomida17}. 
The spiral arm induced by the disk gravitational instability becomes clear (Figure~\ref{f4}) as the disk size increases (Figure~\ref{f6}g).
We can see clear spiral patters in the models with large $\theta_0$ that have smaller $Q_{\rm disk}$ (Figures~\ref{f6}f and l).

Figure~\ref{f8} shows the mass accretion rate and protostellar mass for all models plotted against the elapsed time.
The top panel shows that the time variation of the mass accretion rate is greater in the misalignment models than in the alignment model.
This is consistent with the $\theta_0$-dependence of the disk gravitational instability expected from the time evolution of $Q$ parameter (Figures~\ref{f6}f and l).
The bottom panel indicates that the protostellar mass is greater in the misalignment models than in the alignment model.
Both panels indicate that the mass accretion rate tends to be high when the initial angle difference $\theta_0$ is large. 
Figures~\ref{f6} and \ref{f8} show that both the protostellar mass and disk mass are greater in the misalignment models than in the alignment model.

The total mass flowing into the central region is roughly the same among the models, but the alignment model actually has a massive outflow that compensates for the total mass flowing into the central region.
Figure~\ref{f9} shows the outgoing mass plotted against the radial velocity at $t_{\rm ps} = 5000$\,yr for all models in Simulations A.
We calculate masses in every bin of the radial velocity as
\begin{equation}
  m_{\rm out} (\vrad) = \int_{\vrad-\Delta v}^{\vrad+\Delta v} \rho \, dV \ ,
  \label{eq:1}
\end{equation}
and cumulative masses
\begin{equation}
  M_{\rm out} (\vrad) = \int_{\vrad}^{\infty} \rho \, dV \ .
   \label{eq:2}
\end{equation}
The figure indicates that the outflow mass (e.g., with $\vrad \gtrsim 0.2\,\kms$) for the alignment model is about two orders of magnitude greater than that for the perpendicular model.
Thus, a large part of the outgoing mass is removed from the central region by the outflow. 
Figure~\ref{f9} also indicates that the outflow mass decreases with increasing $\theta_0$. 
Our results are consistent with \citet{li13} who claimed that the outflow becomes stronger in the alignment model than the misalignment one.

\subsection{Simulations B: $\theta_0$ and $\alpha_0$ dependences}
\label{sec:SimBresults}

\begin{figure*}
  \begin{center}
    \includegraphics[width=1.0\linewidth]{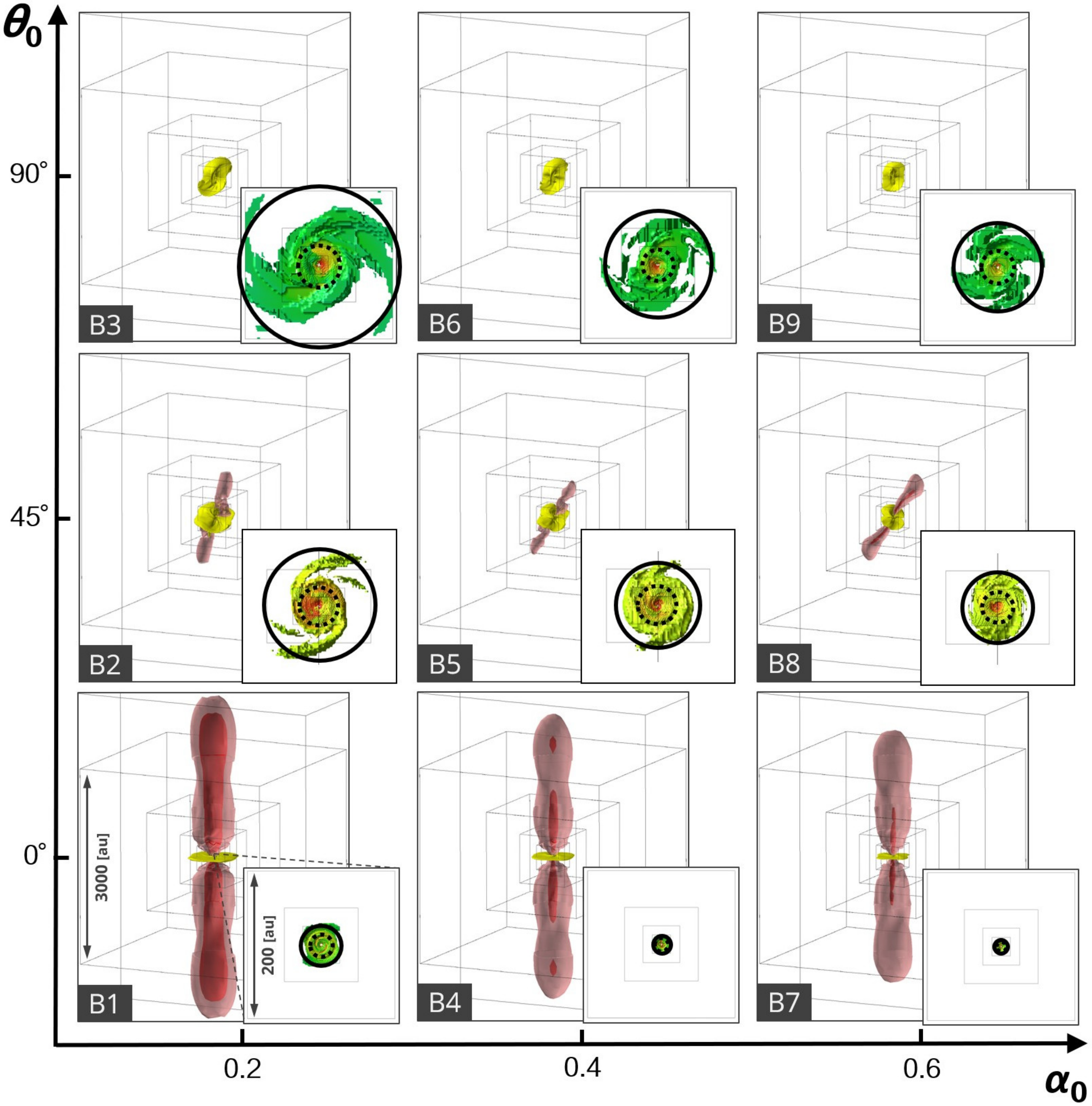}
  \end{center}
\caption{
Three-dimensional view of the outflow, pseudodisk, and rotationally-supported disk at $t_{\rm ps} = 5000$\,yr with a box size of $3000$\,au and $200$\,au (inset) for all models in Simulations B.
In each inserted figure (bottom right corner), the viewing angle is adjusted to become face-on to the disk surface.
Each model (B1--B9) is placed on the parameter space of $\alpha_0$ (horizontal axis) and $\theta_0$ (vertical axis), in which the model name is described in the left bottom corner.
In each panel, the outflow is represented by the isovelocity surfaces of $v_{\rm rad} = 1$ (pink) and $3\,\kms$ (red), respectively.
The pseudodisk is represented by the yellow surface corresponding to an isodensity of $\nh = 5 \times 10^8\,\cc$.
In the inserted figure, the rotationally-supported disk is plotted in colors that represent the density on the disk surface.
The solid and dashed circles are the disk radii $R_{\rm disk}$ and $R_{90\%}$, respectively.
The nested squares in each panel indicate the boundaries of a nested grid. 
}
\label{f10}
\end{figure*}

\begin{figure*}[t]
  \begin{center}
    \includegraphics[width=0.9\linewidth]{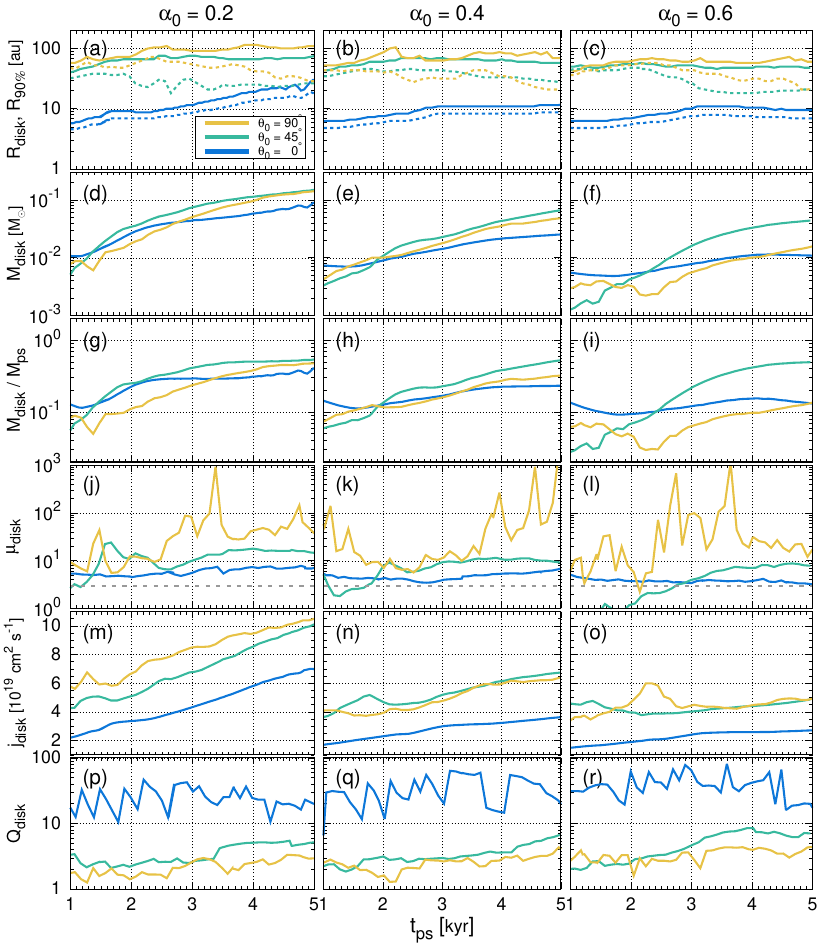}
  \end{center}
\caption{
Time evolution of the disk and protostar properties for all models in Simulations B against the elapsed time during $t_{\rm ps} = 1000$--$5000$\,yr: (panels a, b, c) radii $R_{\rm disk}$ (solid) and $R_{90\%}$ (dotted), (d, e, f) disk mass, (g, h, i) ratio of disk mass to protostellar mass, (j, k, l) magnetization (normalized mass-to-flux ratio), (m, n, o) specific angular momentum, and (p, q, r) Toomre Q parameter averaged over the entire disk.
The dashed line in the panels (j), (k) and (l) plots the initial mass-to-flux ratio, $\mu_0 = 3$.
}
\label{f11}
\end{figure*}

\begin{figure*}[t]
  \begin{center}
    \includegraphics[width=0.9\linewidth]{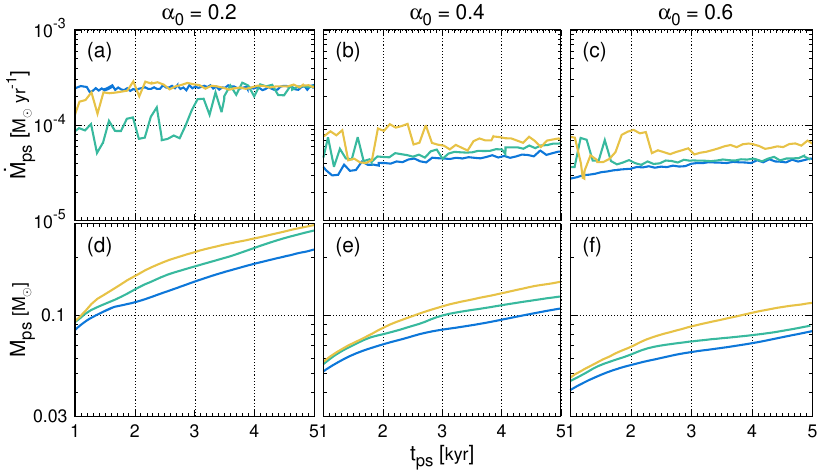}
  \end{center}
\caption{
Mass accretion rate (top panels) and protostellar mass (bottom panels) versus the elapsed time after protostar formation for all models in Simulations B.
}
\label{f12}
\end{figure*}

\begin{figure*}[t]
  \begin{center}
    \includegraphics[width=0.95\linewidth]{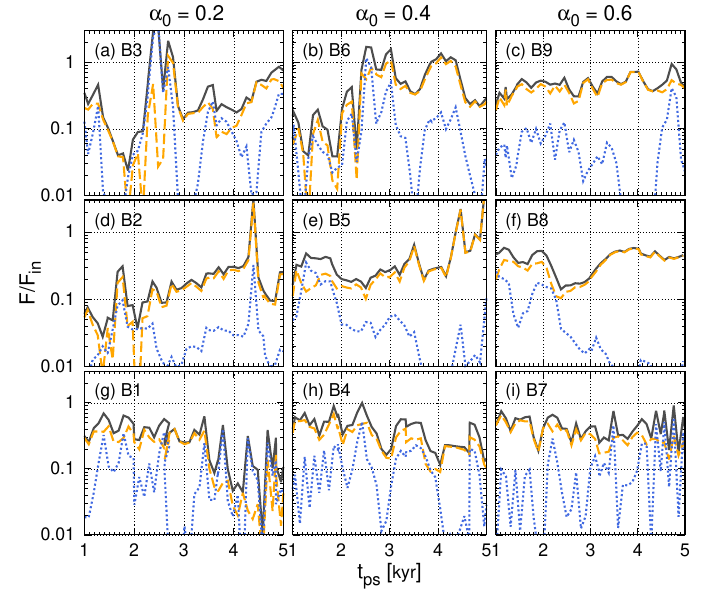}
  \end{center}
\caption{
Same as in Figure~\ref{f7} but for all models in Simulations B.
$(F_{\rm mag} + F_{\rm out})/F_{\rm in}$ (solid lines), $F_{\rm mag}/F_{\rm in}$ (dashed), and $F_{\rm out}/F_{\rm in}$ (dotted) are plotted against the elapsed time after protostar formation.
}
\label{f13}
\end{figure*}

This subsection shows the results of Simulations B. 
\citet{tsukamoto18} pointed out that initial cloud stability, which can be represented by the parameter $\alpha_0$, affects the disk formation. 
When the prestellar cloud is in a highly unstable state, the cloud rapidly collapses while maintaining a nearly spherical configuration.
In such a cloud, the mass accretion rate becomes high \citep{matsushita17}. 
On the other hand, when the prestellar cloud is in a nearly stable state, the gas fluid near the rotation axis, which has a relatively small angular momentum, preferentially falls onto the central region. 
For these reasons, \citet{tsukamoto18} claimed that a relatively large-sized disk forms in the cloud with a small $\alpha_0$.
However, as described above, since \citet{tsukamoto18} only calculated a very early phase of the disk formation, we could not know whether a large-sized disk really forms in a highly unstable state. 
To investigate the effect of the initial cloud stability on the disk formation, we prepared the set of models in Simulations B.

As listed in Table~\ref{t2}, for the models of Simulations B, we changed both the parameters $\theta_0$ and $\alpha_0$ in order to investigate the effects of the angle difference and gravitational stability on disk formation.
Although we adopted various values of $\theta_0$ in Simulations A, we used only three different angles $\theta_0 = 0^\circ$ (alignment model), $45^\circ$ (misalignment), and $90^\circ$ (perpendicular) for Simulations B. 
We also adopted three different values for $\alpha_0$ ($= 0.2$, $0.4$, and $0.6$).
For these models, we also calculated the cloud evolution until $t_{\rm ps} = 5000$\,yr.

The calculation results for all models in Simulations B are plotted in Figure~\ref{f10}.
We can see the emergence of a strong outflow in the alignment models with $\theta_0 = 0^\circ$ (bottom panels), while a weak outflow appears in the misalignment models with $\theta_0 = 45^\circ$ (middle).
On the other hand, the perpendicular models with $\theta_0 = 90^\circ$ (top) did not show a noticeable outflow by the end of the simulation. 
This clearly shows that the outflow strength weakens with increments of $\theta_0$, as shown in \S\ref{sec:SimAresults}.
Figure~\ref{f10} also indicates that the outflow strength depends weakly on $\alpha_0$, in which a relatively strong outflow appears in the model with a smaller $\alpha_0$.
Thus, a strong outflow develops in an initially unstable cloud as pointed out by \citet{matsushita17,matsushita18}.

\begin{figure*}[t]
  \begin{center}
    \includegraphics[width=1.0\linewidth]{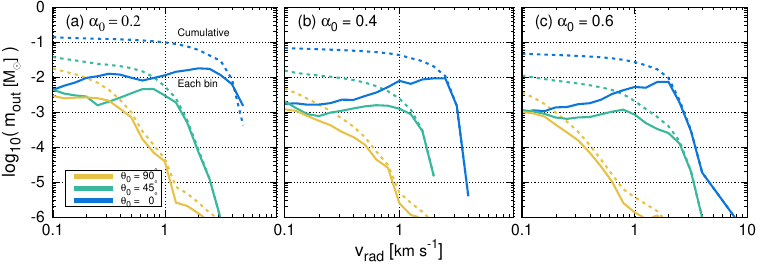}
  \end{center}
\caption{
Mass versus radial velocity at the end of the simulation ($t_{\rm ps} = 5000$\,yr) for (panel a) models B1--B3 (with $\alpha_0 = 0.2$), (b) B4--B6 ($0.4$), and (c) B7--B9 ($0.6$).
Solid and broken lines plot the mass in each velocity bin and the cumulative mass, respectively.
}
\label{f14}
\end{figure*}

In Figure~\ref{f11}, we quantitatively compare the disk properties for the models in Simulations B, as in Figure~\ref{f6}.
The time evolution of each quantity is plotted during $t_{\rm ps} = 1000$--$5000$\,yr. 
Figures~\ref{f11}(a)--(c) plot the time evolution of the disk radii $R_{\rm disk}$ and $R_{90\%}$
 and indicate that the disk size increases as the angle difference $\theta_0$ increases (see also \S\ref{sec:SimAresults}).
The figure also shows that the disk size decreases as the parameter $\alpha_0$ increases, while the difference in the disk size among models with different $\alpha_0$ is not very large.
Model B3, with the smallest $\alpha_0$ ($=0.2$) and the largest $\theta_0$ ($=90^\circ$), has the largest disk, while model B7 with the largest $\alpha_0$ ($=0.6$) and the smallest $\theta_0$ ($=0^\circ$) has the smallest disk.
The disk size at the end of the simulations are plotted as the solid circle in Figure~\ref{f10}.

Figures~\ref{f11}(d)--(f) plot the time evolution of the disk mass. 
Also, the disks in the misalignment and perpendicular models tend to have a greater mass than in the alignment models. 
However, the difference is not very clear, and the disk mass-magnitude relation depends on the evolutionary stage.
For example, at the end of the simulation ($t_{\rm ps} = 5000$\,yr), the model with $\theta_0 = 90^\circ$ has the most massive disk among models with $\alpha_0 = 0.2$, while the disks in the models with $\theta_0 = 45^\circ$ are the most massive in the clouds with $\alpha_0 = 0.4$ and $0.6$.

Figures~\ref{f12}(d)--(f) show that the protostellar mass is greater in the models with a smaller $\alpha_0$ than in the models with larger $\alpha_0$. 
As shown in \citet{matsushita17}, the mass accretion rate is proportional to $\alpha_0^{-3/2}$.
Thus, it is natural that the model with a small $\alpha_0$ has greater protostellar and disk masses, because a large amount of gas falls into the central region for a short duration (Figures~\ref{f12}(a)--(c)). 
In addition, when the parameter $\alpha_0$ is fixed, Figure~\ref{f12}(d)--(f) indicate that the protostellar mass increases with increasing $\theta_0$.
Since a part of the mass flowing to the central region is sent back to the surrounding cloud by the outflow, we need to consider the effect of the outflow when considering the total mass falling into the central region, as described in \S\ref{sec:SimAresults}. 
The parameter $\alpha_0$ regulates the mass and angular momentum flowing to the central region, while the parameter $\theta_0$ determines the efficiency of the outward transport of the angular momentum. 
Since some factors are mixed in the protostar and disk formation process, it is hard to tell which effect is most important for disk formation.

Figures~\ref{f11}(j)--(l) show the magnetization (or the normalized mass-to-flux ratio) of the disk.
In addition to the dependence on $\theta_0$ discussed in \S\ref{sec:SimAresults}, $\mu_{\rm disk}$ becomes slightly lower with increasing $\alpha_0$ throughout the simulation.
This dependence is considered to reflect the difference of the total magnetic flux introduced in the disk.
The accretion rate of both mass and magnetic flux is high in the model with lower $\alpha_0$, which results in a disk with a strong magnetic field, i.e., low mass-to-flux ratio.

The specific angular momentum is a useful index for comparison of disks among models. 
Figures~\ref{f11}(m)--(o) indicate that the specific angular momentum is greater in the misalignment model than in the alignment model.
Independent of $\alpha_0$, there exists a significant different between the alignment and misalignment models.
Figure~\ref{f13} plots the angular momentum fluxes carried away by the magnetic torque and outflows normalized by the flux carried in by the accretion flow.
The ratio of the outgoing flux to the incoming one is slightly lower in the misalignment model than in the alignment model.
Furthermore, the normalized flux increases with increasing $\alpha_0$.
The dependence on these two parameters results in the differences of the specific angular momentum among models (Figures~\ref{f11}m--o).

Figures~\ref{f11}(p)--(r) plot the Toomre parameter $Q_{\rm disk}$.
As seen in models of Simulations A, $Q_{\rm disk}$ decreases as $\theta_0$ increases.
In addition, $Q_{\rm disk}$ slightly decreases as $\alpha_0$ decreases.
For example, Figure~\ref{f10} shows that the spiral arm appears most strongly in model B3.
It is natural that a high mass accretion rate (or small $\alpha_0$) tends to produce a gravitationally unstable disk associated with apparent spiral arms. 
The dependence of the mass accretion rate and protostellar mass on the parameters $\alpha_0$ and $\theta_0$ can be confirmed in Figure~\ref{f12}.

Figure~\ref{f14} shows the outgoing mass plotted against the radial velocity at $t_{\rm ps} = 5000$\,yr.
In addition to the mass in each velocity bin, the cumulative mass is also shown.
The cumulative masses with $v_{\rm rad} \geq 0.1\,\kms$ for the alignment models ($\theta_0 = 0^\circ$) are $M_{\rm out} \sim 0.14$, $0.068$, and $0.046\,\msun$ for $\alpha_0 = 0.2$, $0.4$, and $0.6$, respectively.
The masses for the misalignment models ($\theta_0 = 45^\circ$) are $M_{\rm out} \sim 0.038$, $0.015$, and $0.011\,\msun$, which are several times lesser than those for the alignment models.
In addition, the masses for the perpendicular models ($\theta_0 = 90^\circ$) are $M_{\rm out} \sim 0.018$, $0.0047$, and $0.0037\,\msun$, which are about one orders of magnitude less than those for the alignment models.
Furthermore, the alignment models have the high-velocity component of the outflow ($v_{\rm rad} > 1\,\kms$), while the outflow for the perpendicular models have only the slow-velocity component $v_{\rm rad} \lesssim 1\,\kms$.

The cumulative mass $M_{\rm out} \sim 0.1\,\msun$ is comparable to the disk and protostellar masses for the alignment models.
Thus, the outflow plays a significant role in the disk formation in these cases.
On the other hand, for the perpendicular models, the outflow mass $M_{\rm outflow} \ll 0.01\,\msun$ is much less than the protostellar and disk masses. 
In this case, the outflow does not play a significant role in the disk formation, as described by \citet{li13}.

Figure~\ref{f10} shows that, in addition to $\theta_0$, the initial cloud stability $\alpha_0$ is a significant factor to determine the efficiency of the angular momentum transfer. 
In an initially nearly stable cloud, which has a large $\alpha_0$, the angular momentum is excessively transferred by magnetic braking and/or outflow and a small-sized disk appears.
On the other hand, in an initial unstable cloud with a small $\alpha_0$, a massive disk tends to appear because the gas and angular momentum are advected to the central region in a short duration.

\begin{figure*}[t]
  \begin{center}
    \includegraphics[width=1.0\linewidth]{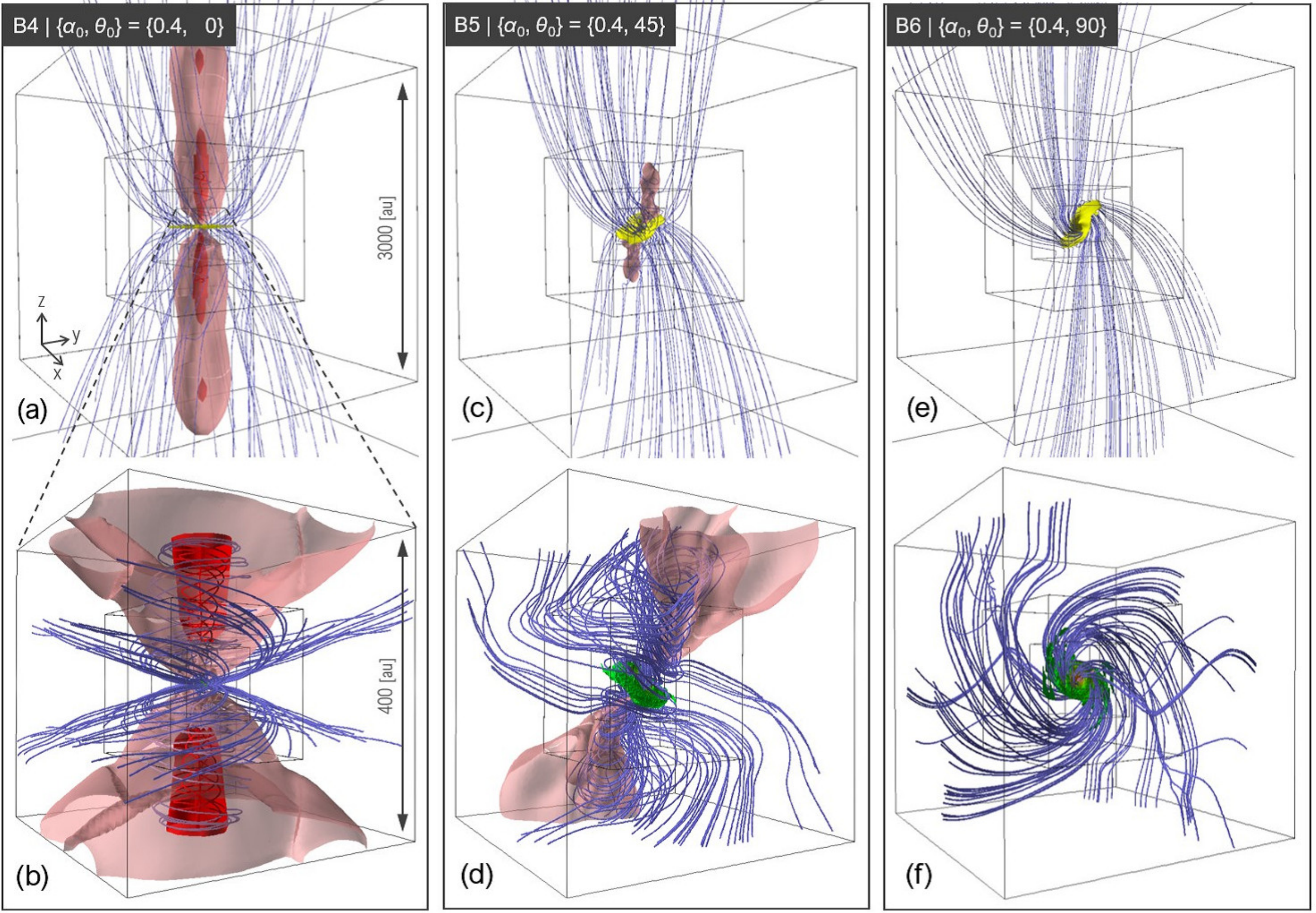}
  \end{center}
\caption{
Three-dimensional view of magnetic field lines (blue streamlines) for models B4 ($\theta_0 = 0^\circ$; panels a and b), B5 ($\theta_0 = 45^\circ$; c and d), and B6 ($\theta_0 = 90^\circ$; e and f) at $t_{\rm ps} = 5000$\,yr.
Outflow (pink and red isovelocity surfaces of $v_{\rm rad} = 1$ and $3\,\kms$, respectively), pseudodisk (yellow isodensity surface of $\nh = 5 \times 10^8\,\cc$), and Keplerian disk (green surface) are also plotted.
The box sizes are $3000$ (top) and $400$\,au (bottom), respectively.
}
\label{f15}
\end{figure*}

\section{Discussion}

In \S\ref{sec:results}, we showed that, in the very early accretion phase (Figure~\ref{f6}a), the rotationally-supported disk is larger in the alignment model ($\theta_0 = 0^\circ$) than in the misalignment models ($\theta_0 \ne 0^\circ$).
However, the disk growth rate is greater in the misalignment models than in the alignment models.
Thus, in the later accretion phase, the disk size in the misalignment models exceeds that in the alignment models (Figures~\ref{f6}e and \ref{f11}a--c). 
At the end of the simulation ($t_{\rm ps} = 5000$\,yr), the disks in the misalignment models are about $3$--$5$ times larger than that in the alignment models.

We can also see a similar trend in the disk mass (Figures~\ref{f6}c, \ref{f6}h, and \ref{f11}g--i). 
The disk is more massive in the alignment models than in the misalignment models in the very early accretion phase, while the disk becomes more massive in the misalignment models than in the alignment models in the later accretion phase.
The disk radius and mass at $t_{\rm ps} = 5000$\,yr for all the models are described in Table~\ref{t1}.
In summary, a relatively large-sized disk is seen in the alignment models only in the early accretion phase, while the disk in the misalignment models grows and its size becomes larger than in the alignment models in the later accretion phase. 
Our simulations showed that the relation between the disk size and mass among models is changed during the mass accretion phase.

\citet{joos12} pointed out that a fan-shaped configuration of magnetic field is finally realized in the alignment case. 
As described in \S\ref{sec:past}, the moment of inertia is larger in the fan-shaped configuration than in the spiral configuration which is realized in the perpendicular case.
Thus, the angular momentum is effectively transferred in the alignment case if the fan-shaped configuration is realized with $\theta_0 = 0^\circ$. 
However, the configuration of the magnetic field will vary with time. 
In the later accretion phase, by which time a large fraction of the cloud mass has fallen onto the central region, a fan-shaped configuration would be realized.
On the other hand, in the very early phase, the configuration of the magnetic field would be represented more nearly by a uniform-parallel configuration as shown in Figure~\ref{f1}.

To confirm the configuration of the magnetic field in the later accretion phase, Figure~\ref{f15} plots magnetic field lines for the alignment model B4 ($\theta_0 = 0^\circ$), the misalignment model B5 ($\theta_0 = 45^\circ$) and the perpendicular model B6 ($\theta_0 = 90^\circ$) at the end of the calculation.
In B4, we can confirm an hourglass structure on the large scale and a fan-shaped configuration on the small scale. 
On the other hand, in B6, a nearly hourglass structure appears on the large scale, while a highly twisted configuration of magnetic field lines can be seen on the small scale.
Thus, we can confirm a rough agreement between schematic view (Figure~\ref{f1}) and magnetic field lines derived from the simulations (Figure~\ref{f15}). 
However, more realistically, the magnetic field lines have a very complicated configuration. 
For example, in the alignment model B4, the poloidal components of magnetic field have a fan-shaped configuration, while strong toroidal components also exist. 
A strong toroidal field would thicken the pseudodisk, as pointed out by \citet{hennebelle09}.
In addition, a very complicated configuration of magnetic field lines is realized in B5. 
For this model, it is very difficult to model the configuration of the magnetic field.\footnote{The magnetic field lines for these models can be seen from various viewing angles at the following link (https://jupiter.geo.kyushu-u.ac.jp/hirano/MHDdisk.html).}

In the aligned case, the magnetic field configuration is changed from a uniform parallel to a fan-shaped configuration as shown in Figure~\ref{f15}.
As described in \S\ref{sec:past}, the magnetic braking is more effective in a fan-shaped configuration than in uniform parallel configuration. 
Thus, the change of the disk size between the alignment and misalignment models during the mass accretion phase is partly owing to the change of the magnetic field configuration. 
However, since the configuration of the magnetic field varies with time, it is very difficult to estimate the moment of inertia and the efficiency of the magnetic braking from the simulations.

In addition to the change in the magnetic configuration, the outflow strength would be related to the disk formation \citep{li13}.
Also, the existence of a flattened pseudodisk may affect the disk evolution \citep{hennebelle09}. 
With these simulations, we confirm that all the factors (initial angle difference, magnetic field configuration, outflow emergence, existence of thick pseudodisk, and gravitational stability) described in past studies are relevant to determining the disk properties. 
We do not identify which mechanism is the most effective because it should depend on the initial conditions of the prestellar clouds such as density and velocity distributions, magnetic field strength, and rotation rate.

\begin{figure*}[t]
  \begin{center}
    \includegraphics[width=1.0\columnwidth]{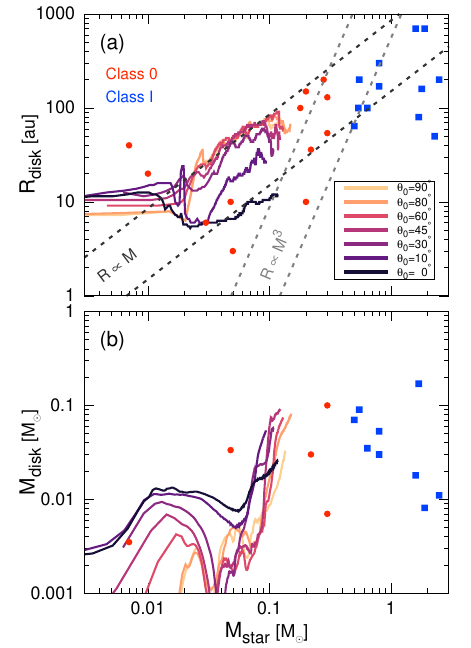}
    \includegraphics[width=1.0\columnwidth]{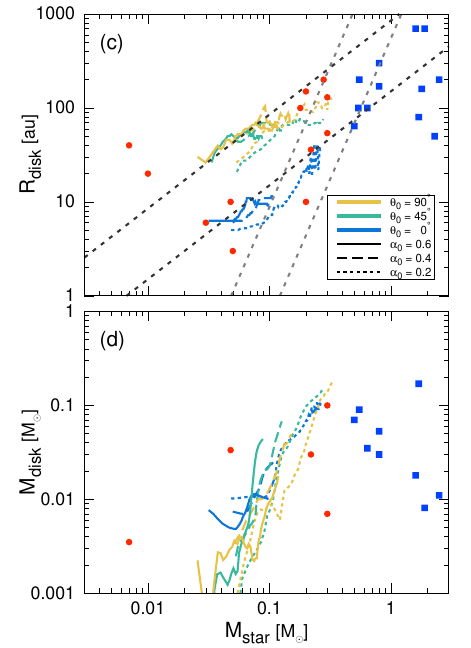}
  \end{center}
\caption{
Time sequence of disk radius (panels a and c) and disk mass (b and d) against the protostellar mass for all models in Simulations A (left) and B (right).
The observation data of Class 0 (red circle) and Class I (blue square) protostars \citep{jorgensen09,yen15,yen17,okoda18} are also plotted.
Black and gray dotted lines in the top panels are models of \citet{terebey84} and \citet{basu98}. 
}
\label{f16}
\end{figure*}

Finally, we comment on the non-ideal MHD effects.
\citet{hennebelle09} and \citet{joos12} claimed that no disk appears when the magnetic field is as strong as $\mu_0 \lesssim 5$.\footnote{\citet{li13} imposed a constant artificial resistivity to some models. However, in their study, since the sink radius is too large to capture the disk (see \S\ref{sec:past}), we do not comment on it here.}
However, with such a strong magnetic field, a Keplerian disk appears and is maintained for $\sim\!5000$\,yr in our study, as in those of \citet{masson16} and \citet{tsukamoto18}.
The difference among the studies is in whether or not non-ideal MHD effects are included.
In the alignment case, a small Keplerian disk can form when a strong magnetic field as well as non-ideal MHD effects are included.

\section{Comparison with observations}

As shown in \S\ref{sec:results}, a Keplerian disk appears in all models.
However, the disk size depends on the parameter $\theta_0$. 
In this section, we compare the disk size derived from simulations with that from observations, in order to verify the simulation results.

Figure~\ref{f16} shows the time sequence of disk radius $R_{\rm disk}$ for all models against protostellar mass $M_{\rm star}$, in which the observed disk radius for Class 0, 0/I, and I protostars are superimposed. 
Although there are some exceptions, the disk radius taken from the simulations are comparable to the observations. 
For models A1--A7 (panel a), we only changed $\theta_0$, while other parameters of the prestellar core such as mass, radius, magnetic field strength and rotation rate are fixed. 
Nevertheless, the disk sizes are in rough agreement with observations. 
Figures~\ref{f16}(c) and (d) indicate that the dependence of the disk size and mass on $\alpha_0$ is weaker than on $\theta_0$.
The variability in disk radius can be explained only by the initial angle difference.
This means that the initial angle difference may be a significant factor in determining the disk properties.
Note that since the protostellar mass is  $\sim0.1\msun$ at the end of our simulations, we cannot compare simulation results with observations in the range of $M_{\rm star}\gtrsim0.1\msun$.
We need a further time integration of simulations to compare with observed data at later stages.
In addition,  we require more samples at earlier stages ($M_{\rm star} \lesssim 0.1\,\msun$) in observations to more reliably compare observations with simulations.

Finally, we comment on the oscillation in the disk radius and mass.
In Figure~\ref{f16}(b), the disk masses have a local peak around $M_{\rm star} \sim 0.01\,\msun$, at which the disk mass $M_{\rm disk}$ is comparable to the protostellar mass $M_{\rm star}$. 
In such a situation, gravitational instability occurs and the gas in the disk rapidly falls onto the central protostar. 
As shown in Figure~\ref{f8}(a), the mass accretion rate shows a high time variability.
After the rapid mass accretion, the disk mass rapidly decreases and the disk radius temporally increases due to the conservation of angular momentum \citep{machida10,tomida17}.
Note that after the rapid mass accretion, the disk mass decreases while the angular momentum remains, leading to the increase of disk radius.
The same phenomenon occurs at $M_{\rm star} \sim 0.1\,\msun$, when the disk mass again becomes comparable to the protostellar mass. 
Thus, the rapid increase and decrease in the disk radius seen in Figure~\ref{f8} can be explained by the gravitational instability (Figure~\ref{f6}l). 
Although we do not further model the gravitational instability, which is outside the scope of this study \citep[for details, also see][]{machida14,machida19,hirano19}, we see that the disk size can be changed in a short duration.

\section{Summary}

In order to resolve the long-standing debate about misalignment, we investigated the star formation process in clouds with the rotation axis inclined against the global magnetic field using 3D non-ideal (resistive) MHD simulations.
We obtained the following results:
\begin{itemize}
\item 
In the very early accretion phase, the disk is larger and more massive in the alignment model ($\theta_0 = 0^\circ$) than in the misalignment models ($\theta_0 \ne 0^\circ$), which is consistent with \citet{matsumoto04} and \cite{tsukamoto18}.
On the other hand, in the later accretion phase, the disk radius and mass in the alignment model are the smallest among the models, which is consistent with \citet{hennebelle09} and \citet{joos12}. 
\item 
The configuration of magnetic field gradually changes in the accretion phase.
In the later accretion phase, the magnetic field has a fan-like configuration in the alignment model, while it has a spiral configuration in the perpendicular model ($\theta_0 = 90^\circ$).
The time variation of magnetic configuration changes the efficiency of magnetic braking, which results in the change of the disk properties. 
\item 
As time proceeds the disk radius and mass in the alignment model becomes relatively small, while those in the misalignment models becomes relatively large.
In the later accretion phase, the disk radius and mass in the misalignment models are several times larger than those in the alignment model. 
\item 
The outflow is weaker in the misalignment models than in the alignment model, as seen in \citet{li13} and \citet{lewis15}. 
A very weak outflow appears in the perpendicular case.
\item 
In the misalignment models, the outflow direction does not agree with that of the global magnetic field because the outflow emerges in the disk normal direction, which roughly corresponds to the rotation vector of the initial cloud. 
The misalignment between outflow and global magnetic field seen in the simulations can provide a reasonable explanation for some observations \citep{hull13}.
\item 
The gravitational stability (or the ratio of thermal to gravitational energy) of the initial cloud also affects the disk formation and outflow driving. 
A large-sized disk and strong outflow tend to appear in an initially unstable cloud, while a small-sized disk and a weak outflow are seen in an initial nearly stable cloud. 
In addition, the initial angle difference between the rotation axis and magnetic field does not significantly affect the disk evolution and outflow driving when the initial cloud is in a highly unstable state.
\end{itemize}

\begin{figure}[t]
  \begin{center}
    \includegraphics[width=1.0\columnwidth]{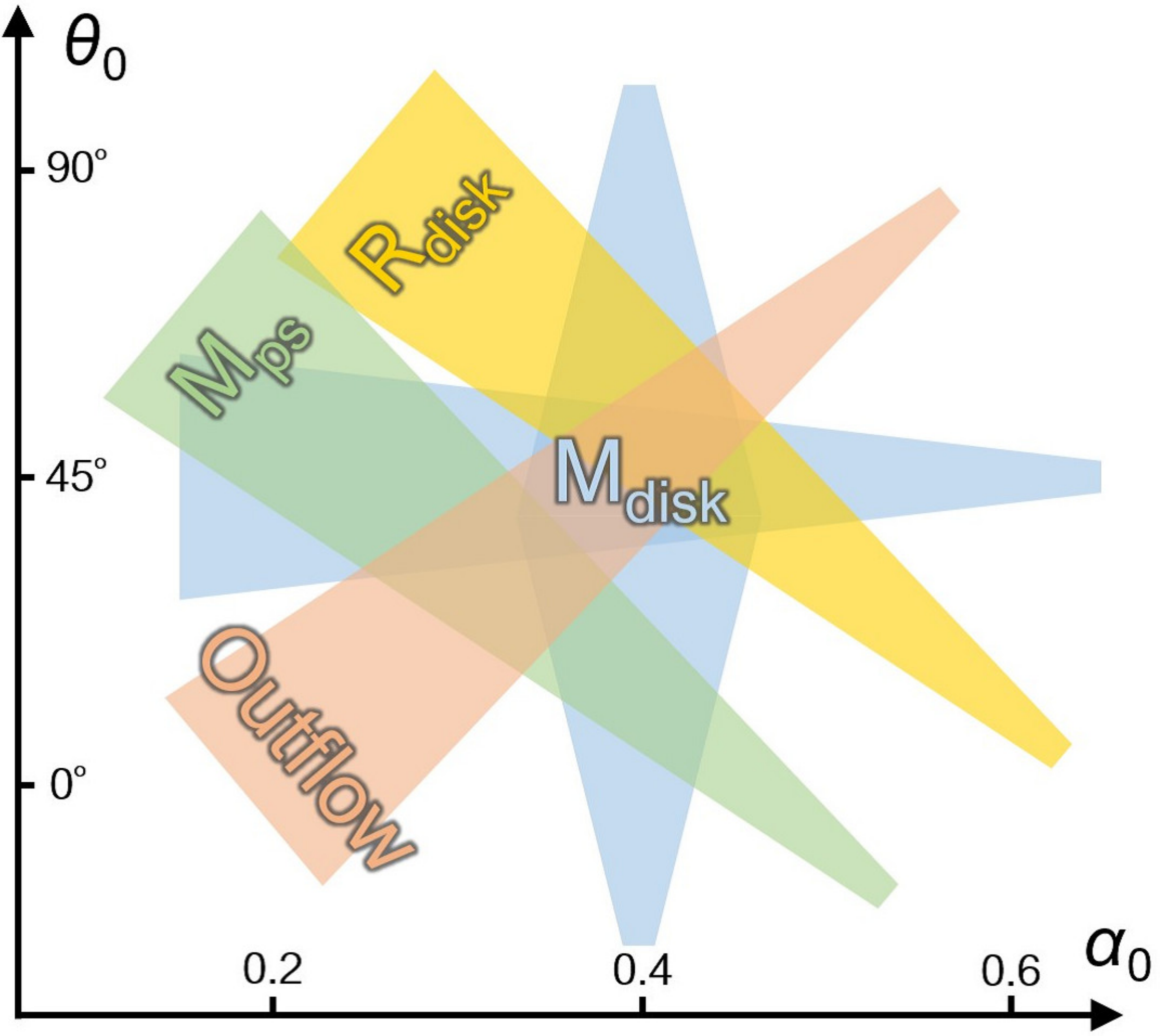}
  \end{center}
\caption{
Schematic view of disk size (yellow), disk mass (blue), protostellar mass (green), and outflow strength (orange) in the parameter space, $\alpha_0$ and $\theta_0$.
The properties are large in the wide region, while they are small in the narrow region. 
Each value is described in Table~\ref{t1}.
}
\label{f17}
\end{figure}

Our simulations showed that the misalignment promotes the disk formation and suppresses the outflow driving in the gas accretion phase.
Figure~\ref{f17} shows that tendency of the disk properties and outflow strength. 
The figure indicates that a massive and large-sized disk tends to appear in the cloud with a large $\theta_0$ and a small $\alpha_0$, while a less massive and small-sized disk appears with a small $\theta_0$ and a large $\alpha_0$.\footnote{
The dependence of the disk mass on $\theta_0$} is not very simple and the most massive disk appears around $\theta_0 = 45^\circ$ at the end of the simulation ($t_{\rm ps} = 5000$\,yr). However, the difference in the disk mass among misaligned models ($\theta_0 \ne 0^\circ$) is very small and the disk mass shows a high-time variability.
Thus, the tendency is expected to be changed in further time integration.
The tendency of the protostellar mass is almost the same as in the disk mass and radius: a massive protostar appears with a large $\theta_0$ and a small $\alpha_0$ and vice versa.
However, it is not clear whether or not the parameter dependence shown in this study is maintained even in the later accretion phase.
Further long-term calculations are necessary to make a conclusion.

In our simulation, unlike \citet{hennebelle09} and \citet{joos12}, a Keplerian disk appears even in the alignment case, although the disk size in the alignment case is significantly smaller than in the misalignment case. 
The difference between our study and \citet{hennebelle09} and \citet{joos12} is the inclusion of non-ideal MHD effect. 
In reality, the formation of a rotationally-supported disk in the alignment case was confirmed in recent non-ideal MHD simulations \citep{masson16,tsukamoto18}. 
Also, the outflow appears even in the misalignment case, while the outflow in the perpendicular case is much weaker than in other cases. 
Thus, although the misalignment nature quantitatively changes the properties of disk and outflow, it does not qualitatively change the star formation process.

\acknowledgments

We have benefited greatly from discussions with Kengo Tomida.
We also thank our anonymous referee for constructive comments on this study.
This work used the computational resources of the HPCI system provided by the supercomputer system SX-ACE at Cyber Sciencecenter, Tohoku University and Cybermedia Center, Osaka University through the HPCI System Research Project (Project ID: hp160079, hp170047, hp180001, and hp190035).
Simulations were also performed by SX-ACE on Earth Simulator at JAMSTEC provided by 2018 and 2019 Koubo Kadai.
This work was supported (in part) by JSPS Research Fellow to SH and JSPS KAKENHI Grant Numbers 18J01296 to SH and by 17K05387, JP17H02869, JP17H06360, 17KK0096 to MNM, and a University Research Support Grant from the National Astronomical Observatory of Japan (NAOJ).
SB was supported by a Discovery Grant from the Natural Sciences and Engineering Research Council of Canada.

\appendix
\restartappendixnumbering

\section{Numerical convergence}
\label{sec:app1}

We ran additional sets of simulations in order to test the numerical convergence of the results. 
While we used $64^3$ cells in each refinement level for the models in the main text, we re-ran models A1, A4, and A7 with higher resolution grids, $128^3$ cells in each refinement level.
We stopped the runs at $t_{\rm ps} = 1000$\,yr after the protostar formation because of the higher computational cost.

\begin{figure}[t]
  \begin{center}
    \includegraphics[width=1.0\columnwidth]{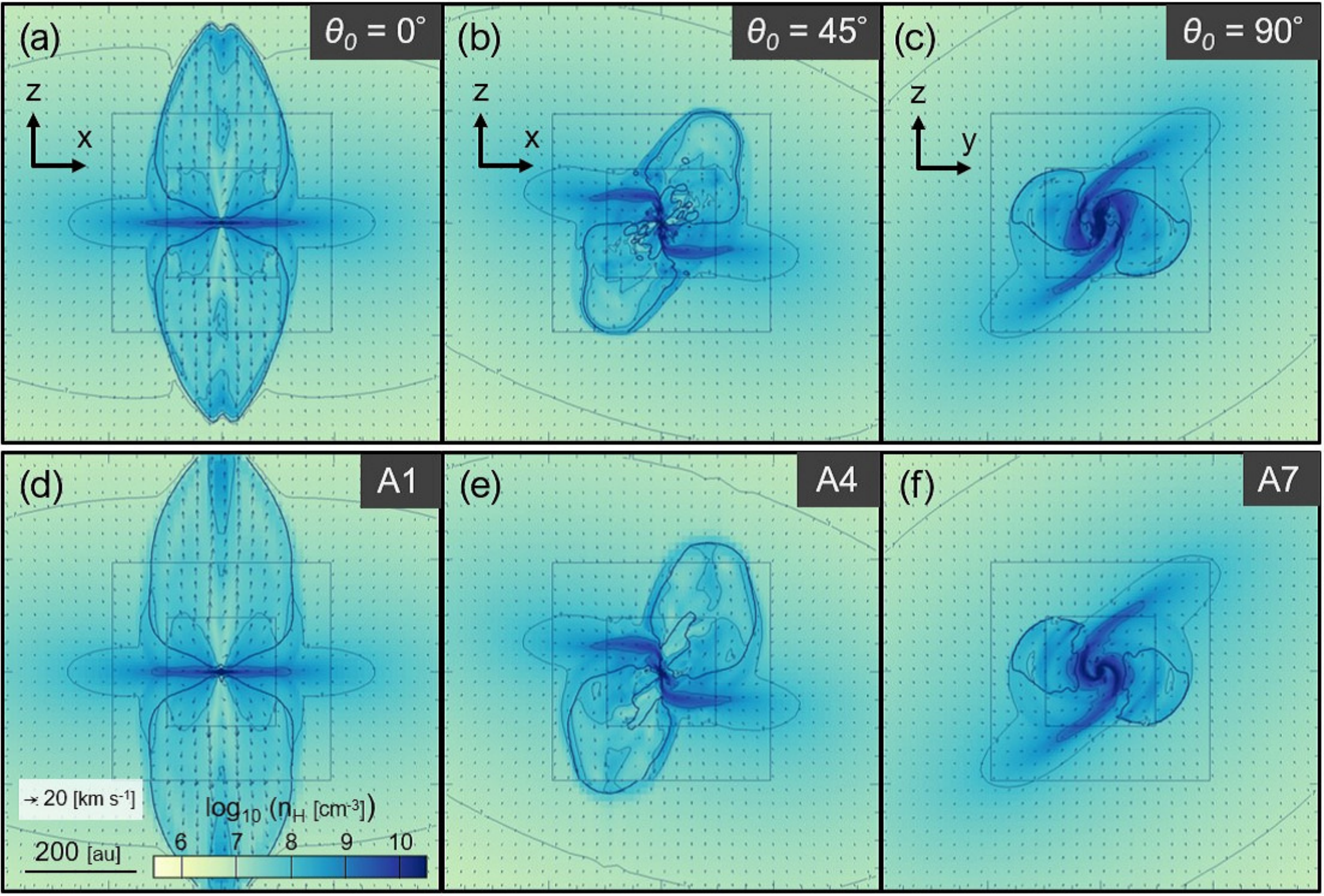}
  \end{center}
\caption{
Same as in Figure~\ref{f2}, but for models A1, A4, and A7 with higher resolution grids, with $128^3$ cells in each refinement level (top panels), and with standard grids, with $64^3$ cells (bottom panels), respectively, at $t_{\rm ps} = 1000$\,yr after protostar formation.
The two panels on the left and the right panel show the $y = 0$ and $x = 0$ planes, respectively.
}
\label{fA1}
\end{figure}

\begin{figure}[t]
  \begin{center}
    \includegraphics[width=0.9\columnwidth]{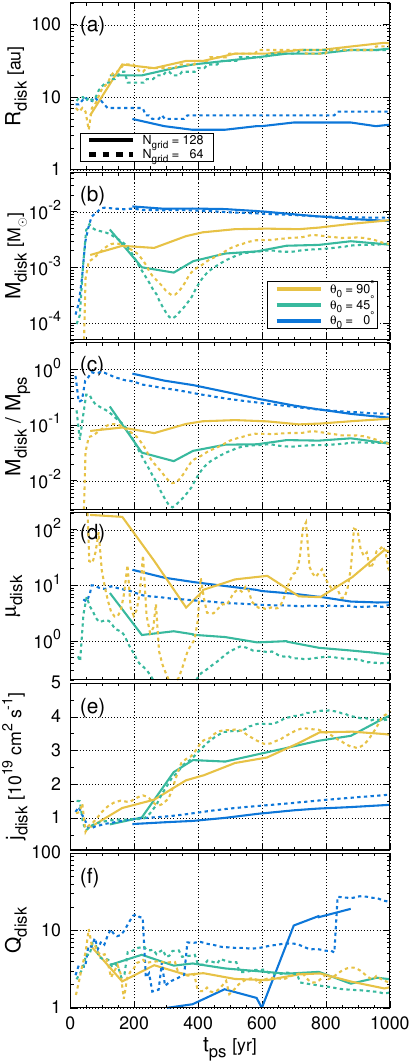}
  \end{center}
\caption{
Same as in Figure~\ref{f6}, but with higher resolution grid models A1, A4, and A7.
The higher resolution grid models ($128^3$ cells in each refinement level) are plotted by solid lines, while the standard grids models ($64^3$ cells) are by the dashed line. The calculations were executed until $t_{\rm ps} = 1000$\,yr after protostar formation
}
\label{fA2}
\end{figure}

Figure~\ref{fA1} shows the density distributions at $t_{\rm ps} = 1000$\,yr.
The effect of different numerical resolutions can be found by comparing the upper ($128^3$ cells) and lower ($64^3$ cells) panels.
There is no significant difference in the overall structure of the rotation disk and expanding outflow.
Figure~\ref{fA2} summarizes the analyzed disk quantities.
In the aligned models (A1), the disk quantities in the higher resolution case show the similar track to the standard case. 
In the misaligned and perpendicular cases (A4 and A7), there are some differences in the disk quantities in the early phase ($t_{\rm ps} < 400$\,yr). 
However, the differences becomes small as time proceeds.  
Furthermore, the dependence of physical quantities on the misalignment parameter ($\theta_0$) are qualitatively the same between the two resolution cases, as shown in each panel of Figure~\ref{fA2}.
Thus, we conclude that the simulation results in this study are qualitatively not significantly affected by the numerical resolution.

\section{Symmetry breaking in perpendicular models}
\label{sec:app2}

The initial clouds adopted in this study have a spatial symmetry such as point symmetry to the origin for the misaligned models and both mirror symmetry to the $y=0$ plane and point symmetry to the origin for the perpendicular models (see Figure~\ref{f1}).
Such symmetries should be maintained during the calculation. 
We can confirm  symmetric structures in the aligned and misaligned models in  Figure~\ref{f2}. 
However, the symmetry is somewhat broken in the perpendicular models due to the numerical perturbation, which may generate a harmful effect.

\begin{figure}[t]
  \begin{center}
    \includegraphics[width=1.0\columnwidth]{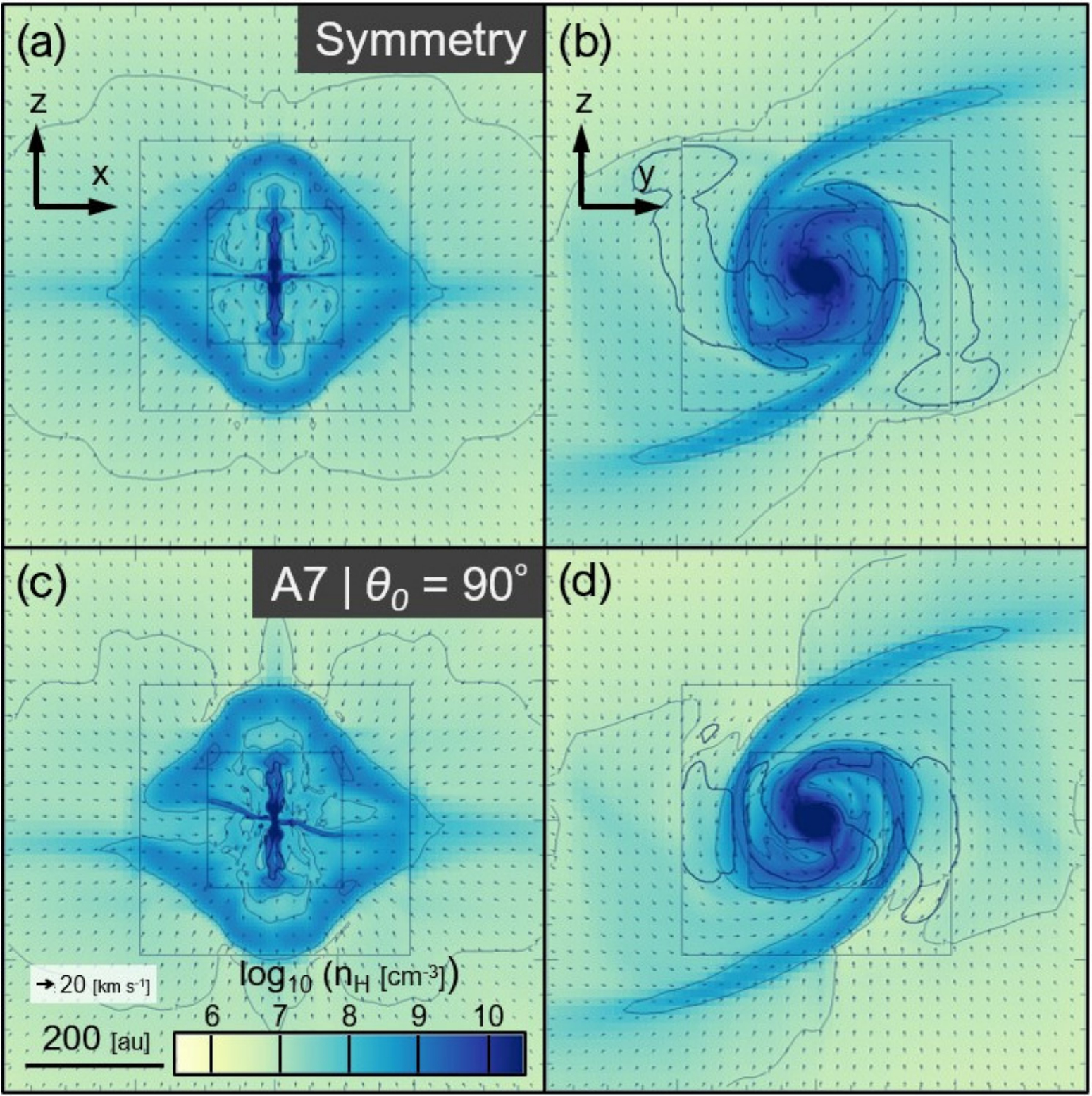}
  \end{center}
\caption{
Same as in Figure~\ref{f2}, but only for models A7 and A7sym ($\theta_0 = 90^\circ$).
The mirror and point symmetries are artificially imposed in the top panels (panels a and b), while no artificial symmetry is imposed in the bottom panels (panels c and d). 
The calculation results at $t_{\rm ps} = 5000$\,yr after protostar formation are plotted on the $y = 0$ (left panels) and $x = 0$ (right panels) planes.
}
\label{fA3}
\end{figure}

\begin{figure}[t]
  \begin{center}
    \includegraphics[width=0.9\columnwidth]{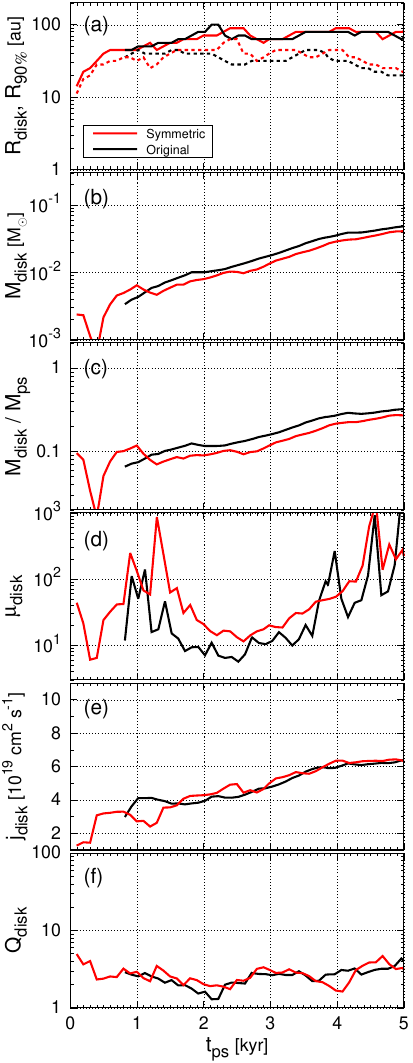}
  \end{center}
\caption{
Same as in Figure~\ref{f6}.
The calculation results for models A7sym (red) and A7 (black, A7).
Each quantity is plotted against the elapsed time after protostar formation.
}
\label{fA4}
\end{figure}

To test whether or not the symmetry breaking affects the simulation results and analysed properties, we performed an additional run for perpendicular model A7,  artificially imposing both the mirror symmetry to the $y=0$ plane and point symmetry to the origin (hereafter we call this model A7sym).
Figure~\ref{fA3}(a) shows the density distribution on the $y = 0$ plane at $t_{\rm ps} = 5000$\,yr after protostar formation for models  A7sym (top panels) and A7 (bottom panels).
The overall structure of model A7sym is very similar to the original run (model A7), in which we did not impose any artificial symmetry in model A7 (Figure~\ref{fA3}c).
Especially, the rotation disks are almost the same regardless of whether or not symmetry breaking occurs (Figures~\ref{fA3}b and d).
Figure~\ref{fA4} shows that the disk properties are almost the same between the two runs.

\end{document}